\documentclass[fleqn,usenatbib]{mnras}

\usepackage[T1]{fontenc}

\DeclareRobustCommand{\VAN}[3]{#2}
\let\VANthebibliography\thebibliography
\def\thebibliography{\DeclareRobustCommand{\VAN}[3]{##3}\VANthebibliography}

\usepackage{graphicx}	
\usepackage{amsmath}	
\usepackage{amssymb}	
\usepackage{tabu}
\usepackage{longtable}
\usepackage{makecell}
\usepackage{multicol,multirow}
\usepackage{xcolor}
\usepackage{ulem}
\usepackage{newtxtext,newtxmath}

\newcolumntype{C}[1]{>{\centering\arraybackslash}p{#1}}

\title[Merger-triggering of radio AGN/Type 2 quasars]{Do AGN triggering mechanisms vary with radio power? II. The importance of mergers as a function of radio power and optical luminosity}

\author[J. C. S. Pierce]{J. C. S. Pierce,$^{1,2}$\thanks{E-mail: j.c.pierce@sheffield.ac.uk}
C. N. Tadhunter,$^{1}$
Y. Gordon,$^{3,4}$
C. Ramos Almeida,$^{5,6}$
S. L. Ellison,$^{7}$
C. O'Dea,$^{3}$
\newauthor L. Grimmett,$^{8}$
L. Makrygianni,$^{9}$
{P. S. Bessiere$^{5}$}
and P. Do\~{n}a Gir\'{o}n$^{5,6}$
\\
\\
$^{1}$Department of Physics and Astronomy, University of Sheffield, Sheffield S3 7RH, UK\\
$^{2}$Centre for Astrophysics Research, University of Hertfordshire, College Lane, Hatfield AL10 9AB, UK\\
$^{3}$Department of Physics and Astronomy, University of Manitoba, Winnipeg, MB R3T 2N2, Canada\\
$^{4}$Department of Physics, University of Wisconsin-Madison, 1150 University Ave, Madison, WI 53706, USA\\
$^{5}$Instituto de Astrof\'{i}sica de Canarias, Calle V\'{i}a L\'{a}ctea, s/n, E-38205 La Laguna, Tenerife, Spain\\
$^{6}$Departamento de Astrof\'{i}sica, Universidad de La Laguna, E-38205 La Laguna, Tenerife, Spain\\
$^{7}$Department of Physics \& Astronomy, University of Victoria, Finnerty Road, Victoria, British Columbia V8P 1A1, Canada \\
$^{8}$Department of Statistics, School of Mathematics, University of Leeds, Leeds LS2 9JT, UK \\
$^{9}$The School of Physics and Astronomy, Tel Aviv University, Tel Aviv 69978, Israel 
}

\date{Accepted XXX. Received YYY; in original form ZZZ}

\pubyear{2021}

\begin{document}
\label{firstpage}
\pagerange{\pageref{firstpage}--\pageref{lastpage}}
\maketitle

\begin{abstract}

Investigation of the triggering mechanisms of radio AGN is important for improving our general understanding of galaxy evolution.
In the first paper in this series, detailed morphological analysis of high-excitation radio galaxies (HERGs) with intermediate radio powers suggested that the importance of triggering via galaxy mergers and interactions increases strongly with AGN radio power and weakly with optical emission-line luminosity. 
Here, we use an online classification interface to expand our morphological analysis to a much larger sample of 155 active galaxies (3CR radio galaxies, radio-intermediate HERGs and Type 2 quasars) that covers a broad range in both 1.4 GHz radio power and [OIII]$\lambda$5007 emission-line luminosity. All active galaxy samples are found to exhibit excesses in their rates of morphological disturbance relative to 378 stellar-mass- and redshift-matched non-active control galaxies classified randomly and blindly alongside them. These excesses are highest for the 3CR HERGs (4.7\,$\sigma$) and Type 2 quasar hosts (3.7\,$\sigma$), supporting the idea that galaxy mergers provide the dominant triggering mechanism for these subgroups. 
When the full active galaxy sample is considered, there is clear evidence to suggest that the enhancement in the rate of disturbance relative to the controls increases strongly with [OIII]$\lambda$5007 emission-line luminosity but not with 1.4 GHz radio power. 
Evidence that the dominant AGN host types change from early-type galaxies at high radio powers to late-type galaxies at low radio powers is also found, suggesting that triggering by secular, disk-based processes holds more importance for lower-power radio AGN.

\end{abstract}





\section{Introduction}
\label{sec:intro}

Through the energetic feedback of their intense radiation fields and powerful relativistic jets, active galactic nuclei (AGN) are thought to have a significant influence on the evolution of their host galaxies \citep[see][for reviews]{cat09,fab12,veil20}. These feedback effects are regularly incorporated into semi-analytic models and hydrodynamical simulations of galaxy evolution as a means of regulating and suppressing star formation, from the scales of galaxy nuclei to the circumgalactic environment \citep[][]{sd15}.

The jets of radio AGN are thought to be particularly important in the feedback context. In galaxy halos and dense large-scale environments (groups and clusters; on scales $\sim$10 kpc -- 1 Mpc), they are thought to regulate the cooling of hot gas, preventing the build up of mass in the most massive galaxies and explaining the decline at the upper end of observed galaxy luminosity and stellar mass functions \citep[][]{bow06,cro06,cro16,vog14}. This type of feedback, often referred to as ``jet mode" or ``maintenance mode", is usually linked with AGN fuelled by low Eddington rate, radiatively-inefficient supermassive black hole (SMBH) accretion -- low-excitation radio galaxies (LERGs), identified through their weak optical emission-line spectra \citep[e.g.][]{but10,bh12,hb14}.

The alternative type of feedback in this widely used scheme is ``quasar mode" or ``radiative mode", which is linked with AGN-driven outflows that influence star formation on the scales of galaxy nuclei ($\sim$1--10 kpc). While this has traditionally been associated with radiatively-driven winds in luminous AGN \citep[e.g.][]{kp15}, there is strong observational evidence to suggest that jets are also capable of driving multiphase outflows \citep[e.g.][]{mor05,mor13,mor15,mah16,huse19a,oost19,sant20} or entraining molecular gas \citep[e.g.][]{mcn14,rus17,rus19} on these scales, even in quasar-like AGN \citep[][]{wm18,jar19,oost19}. Radio AGN that are fuelled by high Eddington rate, radiatively-efficient SMBH accretion -- high-excitation radio galaxies (HERGs) or jetted quasars, which exhibit strong optical emission-line spectra -- are therefore particularly interesting objects in the feedback scenario.

A key aspect for the correct implementation of radio AGN into models of galaxy evolution lies in determining how they are triggered. 
Deep, ground-based optical imaging studies of powerful radio galaxies (L$_{\rm 1.4GHz}$ $\gtrsim 10^{25}$ W\,Hz$^{-1}$) have revealed high rates of morphological disturbance from galaxy mergers and interactions \citep{heck86,sh89a,sh89b,ram11}, events which efficiently transport material to galaxy centres and could provide the dominant trigger for the nuclear activity \citep[e.g.][]{bh96,gabor16}. The merger rates are particularly high for samples of radio AGN with strong optical emission lines, with 94$^{+2}_{-7}$ per cent\footnote{The uncertainties from our previous studies are here updated to binomial $1\sigma$ confidence intervals from the Bayesian technique of \cite{cam11}.} of strong-line radio galaxies (SLRGs\footnote{SLRGs and WLRGs are selected based on [OIII] emission-line equivalent width (EW$_{\rm [OIII]} > 10$\AA{} and EW$_{\rm [OIII]} < 10$\AA{}, respectively), but show a strong overlap with the HERG and LERG populations \citep[see discussion in][]{tad16}.}) in the 2Jy sample exhibiting clear tidal features \citep{ram11}. The 2Jy SLRGs also show an excess of high-surface-brightness tidal features relative to non-active elliptical galaxies matched to the targets in terms of absolute magnitude \citep[][]{ram12}, and are found to preferentially lie in group-like environments that are well suited to frequent galaxy interactions \citep{ram13}.

On the other hand, evidence for interactions in powerful radio galaxies with radiatively-inefficient AGN is much less frequent, with only 27$^{+16}_{-9}$ per cent of 2Jy weak-line radio galaxies (WLRGs$^2$) showing clear tidal features \citep{ram11}. In addition, analysis at both optical \citep[][]{ram13,sab13} and X-ray wavelengths \citep{ine13,ine15} shows that these objects preferentially lie in dense, cluster-like environments, where high relative galaxy velocities can reduce the merger rate \citep[][]{pb06}. It has been proposed that radiatively-inefficient nuclear activity is instead predominantly fuelled by the prevalent hot gas content that is present in these environments, through direct accretion \citep[e.g.][]{allen06,hard07}, cooling flows \citep[e.g.][]{tad89,baum92,best05b}, or the chaotic accretion of condensing cold gas \citep[][]{gas13,gas15,tremblay16}. 

This current picture of the triggering and fuelling of radio AGN is, however, largely based on samples with the highest radio powers, despite the fact that local radio luminosity functions are found to increase steeply towards lower radio powers \citep[][]{ms07,bh12,sad14,sab19}. In addition, the subpopulation of active galaxies that have strong optical emission lines and intermediate radio powers (HERGs and jetted Type 2 quasars with 22.5 $<$ log(L$_{\rm 1.4GHz}$) $< 25.0$ W\,Hz$^{-1}$) show strong evidence for jet-driven multiphase outflows \citep[e.g.][]{tad14b,har15,vm17,jar19}. The broadest ionised gas emission-line profiles for radio AGN in the Sloan Digital Sky Survey (SDSS) are also found to be associated with those in the intermediate radio power regime \citep[][]{mul13}. Study of intermediate-radio-power, radiatively-efficient AGN is thus particularly important for improving our understanding of the role of radio AGN feedback in galaxy evolution. 

In the first paper in this series (\citeauthor{pierce19} \citeyear{pierce19}; hereafter Paper I), we used detailed analysis of the optical morphologies of a sample of 30 local HERGs with intermediate radio powers ($z<0.1$; 22.5 $<$ log(L$_{\rm 1.4GHz}$) $< 24.0$ W\,Hz$^{-1}$) to investigate the importance of merger-based triggering in this subpopulation for the first time. When compared with the 2Jy HERGs and quasars, the rates of disturbance in the higher and lower radio power halves of the sample (67$^{+10}_{-14}$ per cent and 40$^{+13}_{-11}$ per cent, respectively) suggested that interactions have decreasing importance for triggering radio AGN towards lower radio powers. However, a relationship with decreasing optical emission-line luminosity could not be ruled out, and it is interesting to note that the hosts of Type 2 quasars with low-to-intermediate radio powers frequently show clear tidal features in deep, ground-based imaging observations \citep{bess12}. The radio-intermediate HERGs were also shown to exhibit a mixture of morphological types, consistent with the idea that the dominant hosts of radiatively-efficient radio AGN change from massive, early-type galaxies at high radio powers \citep[as shown by e.g.][]{mms64} to late-type galaxies at lower radio powers \citep[like Seyfert galaxies; e.g.][]{adams77}.

Here, we use an online interface to expand our morphological analysis to a much larger sample of 155 local active galaxies covering a broad range in both 1.4 GHz radio power and [OIII]$\lambda$5007 emission-line luminosity \citep[a proxy for the total AGN power; e.g.][]{heck04}, allowing us to investigate the relationships suggested by the results of Paper I in more detail. The new sample includes high-radio-power 3CR HERGs and LERGs, radio-intermediate HERGs, and Type 2 quasar hosts, imaged using an identical observing setup at a consistent limiting surface brightness depth ($\mu_r$$\sim $27 mag\,arcsec$^{-2}$). The inclusion of both radiatively-efficient and -inefficient radio AGN in the 3CR sample also allows the apparent dichotomy in their dominant triggering and fuelling mechanisms at high radio powers to be tested. Crucially, our observations also allow us to select a large control sample of non-active galaxies matched in stellar mass and redshift from the wide image fields, the morphologies of which were classified blindly alongside those of the targets. 

The paper is structured as follows. Details on the samples, observations and image reduction are provided in \S\ref{sec:samp_sel_obs_red}. The online interface used to obtain the morphological classifications is described in \S\ref{sec:methods}, along with the method used to select non-active control galaxies from the target image fields. \S\ref{sec:res} outlines the analysis of the morphological classifications and the subsequent results relating to the rates of disturbance (\S\ref{subsec:dist_rates}), the interaction/merger stage  for disturbed galaxies (\S\ref{subsec:features_merger_stage}), and the galaxy morphological types (\S\ref{subsec:morph_types}). A discussion of both the classification methodology and the results is presented in \S\ref{sec:disc}. The study is then summarised in \S\ref{sec:summary}.
A cosmology described by $H_{0} = 73.0$ km\,s$^{-1}$\,Mpc$^{-1}$, $\Omega_{\rm m} = 0.27$ and $\Omega_{\rm \Lambda} = 0.73$ is assumed throughout this paper, for consistency with our previous work.

\section{Sample selection, observations and reduction}
\label{sec:samp_sel_obs_red}

The samples used in this study comprise powerful 3CR radio galaxies, radio-intermediate HERGs and the hosts of Type 2 quasars, which together encompass a broad range in both 1.4 GHz radio power and [OIII]$\lambda$5007 emission-line luminosity. All objects were observed with the Wide-Field Camera (WFC) on the 2.5m Isaac Newton Telescope (INT) at the Observatorio del Roque de los Muchachos, La Palma, using a consistent observing technique. A summary of the redshift, stellar mass, 1.4 GHz radio power and [OIII]$\lambda$5007 emission-line luminosity properties for each of the four active galaxy samples studied is provided in Table~\ref{tab:samples_summary}.

\begin{table}
\centering
\small
\caption{A summary of the redshift, stellar mass, 1.4 GHz radio power and [OIII]$\lambda$5007 emission-line luminosity properties for the four active galaxy samples studied. The ranges, medians, and standard deviations are presented in the top, middle, and bottom rows, respectively.} 
\label{tab:samples_summary}
\begin{tabular}{C{1.15cm}cccc}
\hline 
              & \makecell{RI-HERG\\low}  & \makecell{RI-HERG\\high} & \makecell{3CR}         & \makecell{Type 2\\quasars} \\  \hline 
\textit{}           & 0.031-0.098 & 0.051-0.149 & 0.050-0.296 & 0.051-0.139   \\
\textit{z}        & 0.074        & 0.110        & 0.174        & 0.111          \\
\textit{}      & 0.019        & 0.032        & 0.077        & 0.025          \\  \hline 
\multirow{3}{*}{$\rm log(M_{*}/M_{\odot}$)}  & 10.0-11.4   & 10.7-11.4   & 10.5-12.7   & 10.6-11.4     \\
                  & 10.7         & 11.2         & 11.4         & 11.0           \\
                   & 0.3          & 0.2          & 0.4          & 0.2            \\ \hline  
\multirow{3}{*}{\makecell{log(L$_{\rm 1.4GHz}$)\\(W\,Hz$^{-1}$)}}    & 22.52-23.98 & 24.01-24.88  & 24.67-28.15 & 22.47-26.22   \\
                   & 23.06        & 24.54        & 26.40        & 23.34          \\
                  & 0.39         & 0.24         & 0.58         & 0.86           \\ \hline 
\multirow{3}{*}{\makecell{log(L$_{\rm[OIII]}$)\\(W)}}    & 32.86-34.24 & 33.00-35.06  & 32.67-36.09 & 35.04-35.86   \\
                    & 33.56        & 33.98        & 34.64        & 35.19          \\ 
                   & 0.35         & 0.46         & 0.80         & 0.19     \\  \hline             
\end{tabular}
\end{table}

\subsection{Radio-intermediate HERGs}
\label{subsec:RI-HERG_sel}

\begin{table*}
	\centering
	\caption{\small Basic information for the 28 targets in the RI-HERG high sample described in \S\ref{subsubsec:RI-HERG_high_sel}. Full table available as supplementary material.} \label{tab:RI-HERG_high_target_info} 
	\begin{tabular}{lcccccccc}
	\hline
	SDSS ID (Abbr.) & $z$ & \makecell{$A_{\rm r}$\\(mag)} & \makecell{SDSS \textit{r} mag\\(mag)} & \makecell{log(L$_{\rm 1.4GHz}$)\\(W\,Hz$^{-1}$)} & \makecell{log(L$_{\rm [OIII]}$)\\(W)} & \makecell{log(M$_{*}$)\\(M$_{\odot}$)} & Obs. date & \makecell{Seeing \small{FWHM}\\ (arcsec)}\\
	\hline
J075244.19$+$455657.4 (J0752+45) & 0.052 & 0.23 & 14.24 & 24.52 & 33.68 & 11.3 & 2018-03-11 & { 0.97} \\
J080601.51$+$190614.7 (J0806+19) & 0.098 & 0.11 & 15.46 & 24.55 & 33.93 & 11.4 & 2018-03-11 & { 1.32} \\
J081755.21$+$312827.4 (J0817+31) & 0.124 & 0.12 & 16.81 & 24.39 & 34.26 & 11.0 & 2018-03-11 & { 1.56} \\
J083655.86$+$053242.0 (J0836+05) & 0.099 & 0.10  & 15.49 & 24.09 & 33.78 & 11.3 & 2018-03-13 & { 1.17} \\
J084002.36$+$294902.6 (J0840+29) & 0.065 & 0.15 & 14.66 & 24.73 & 34.33 & 11.2 & 2018-03-11 & { 1.33} \\
... & ... & ... & ... & ... & ... & ... & ... & ... \\ \hline
\end{tabular}
\end{table*}

The radio-intermediate HERGs were selected from the catalogue of 18,\,286 local radio galaxies produced by \citet{bh12}. These were selected as radio-intermediate based on their 1.4 GHz radio powers\footnote{In the range 22.5 $<$ log(L$_{\rm 1.4GHz}$) $< 25.0$ W\,Hz$^{-1}$. Studies of local radio luminosity functions for AGN and star-forming galaxies show that the latter decline steeply above L$_{\rm 1.4 GHz} \sim 10^{23}$ W\,Hz$^{-1}$ \citep[e.g.][]{sad02,bh12}, but since all radio sources in the \cite{bh12} catalogue were additionally identified as AGN using their optical properties, a lower luminosity density limit was considered for the selection.}, but were separated into two samples covering different ranges for the analysis in this paper. The optical spectroscopic data were extracted from the value-added catalogue maintained by the Max Planck Institute for Astrophysics and the Johns Hopkins University (the MPA-JHU value-added catalogue), which provides raw measurements and derived properties (e.g. stellar masses and star-formation rates) based on analysis of the SDSS DR7 spectra\footnote{Available at: \url{https://wwwmpa.mpa-garching.mpg.de/SDSS/DR7/}.}.

\begin{figure}
\centering
    \includegraphics[width=\linewidth]{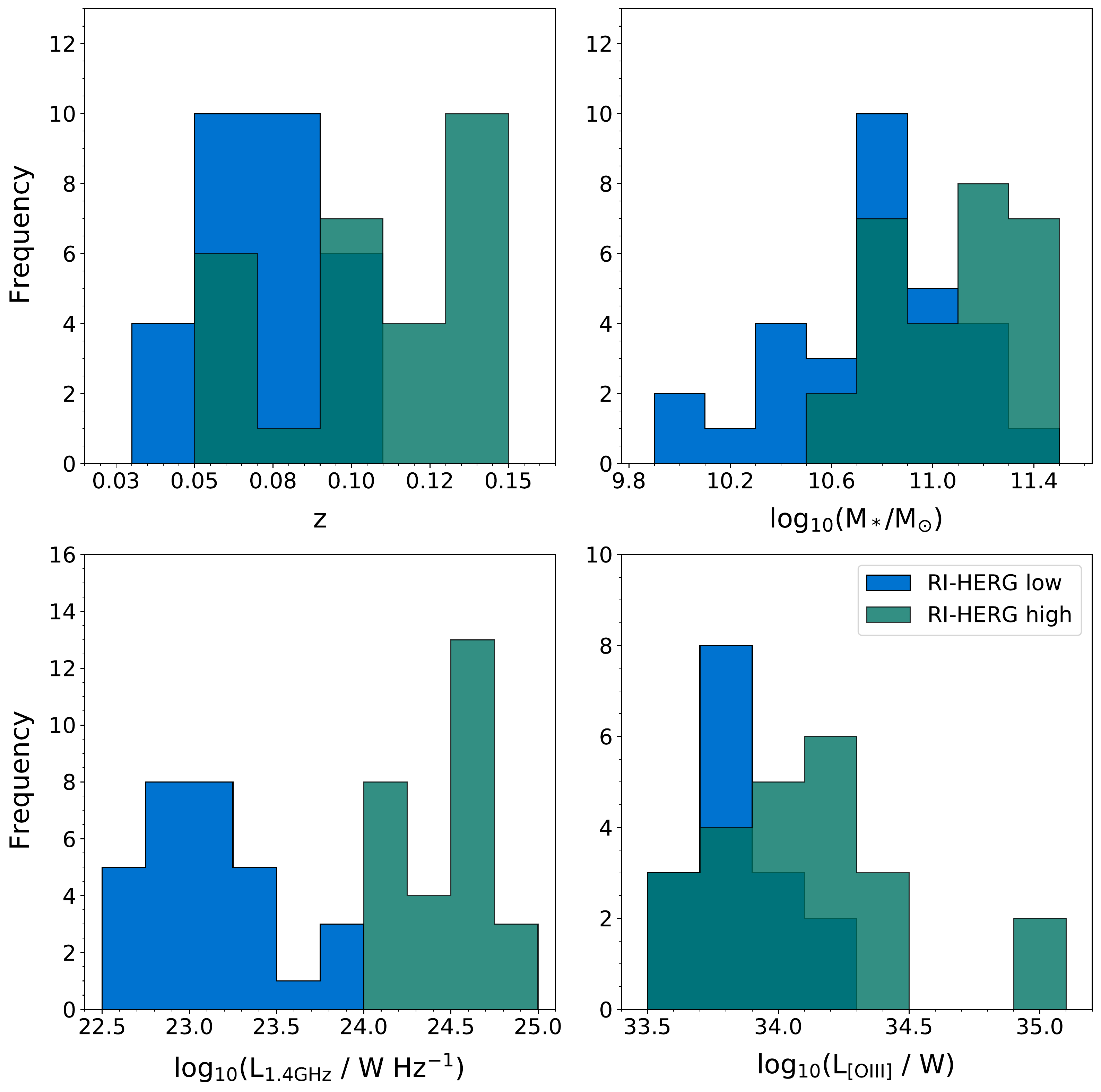}
    \caption{Redshift, stellar mass, 1.4 GHz radio power, and [OIII]$\lambda$5007 emission-line luminosity distributions for the sources in the RI-HERG low (blue) and RI-HERG high (teal) samples.}
   \label{fig:RIHERG_hists}
\end{figure}

\subsubsection{The RI-HERG low sample}
\label{subsubsec:RI-HERG_low_sel}

The detailed optical morphologies of the first of the two radio-intermediate HERG samples were investigated in Paper I. The sample comprises all 30 HERGs in the \citet{bh12} catalogue with radio powers in the range 22.5 $<$ log(L$_{\rm 1.4GHz}$) $< 24.0$ W\,Hz$^{-1}$, redshifts $z < 0.1$ and right ascension coordinates ($\alpha$) in the range 07h 15m $< \alpha <$ 16h 45m. These objects were included in the current project to obtain morphological classifications in a manner consistent with the other active galaxy samples studied in this work, so as to allow direct comparison between the results. Their inclusion also allowed the online interface classification method to be compared with the previous detailed approach (see \S\ref{subsec:method_disc}).
Given the lower radio powers of the objects relative to those in the other radio-intermediate HERG sample studied in this work (\S\ref{subsubsec:RI-HERG_high_sel}), this is referred to as the \textit{RI-HERG low} sample from this point forward. Information on the properties of these objects is provided in Paper I and is not included here, for brevity. 

\subsubsection{The RI-HERG high sample}
\label{subsubsec:RI-HERG_high_sel}

The second radio-intermediate HERG sample selected from \cite{bh12} consists of all HERGs with radio powers in the range 24.0 $<$ log(L$_{\rm 1.4GHz}$) $< 25.0$ W\,Hz$^{-1}$, redshifts $z<0.15$ and right ascension coordinates in the range 07h 45m $< \alpha <$ 15h 45m. Inspection of Faint Images of the Radio Sky at Twenty-cm \citep[FIRST;][]{bec95} images for all objects in this sample (limiting angular resolution $\sim$5") suggested that the radio emission detected by the NRAO VLA Sky Survey \citep[NVSS;][]{con98} would not be contaminated by other radio sources within its $\sim$45" angular resolution limit. NVSS flux densities were therefore used to derive the radio powers in all cases, due to their better sensitivity to extended, low surface brightness emission. These selection criteria resulted in a final sample of 28 radio-intermediate HERGs that is complete within these constraints. 
Some basic properties of these 28 targets are provided in Table~\ref{tab:RI-HERG_high_target_info}, and the distributions of redshift, stellar mass, 1.4 GHz radio power, and [OIII]$\lambda$5007 emission-line luminosity are presented alongside those of the RI-HERG low sample in Figure~\ref{fig:RIHERG_hists}.

\subsection{The 3CR sample}
\label{subsec:3CR_sel}

Our sample of high-radio-power AGN was selected from the \cite{spin85} catalogue of 3CR radio galaxies. The full sample comprises all 84 objects with redshifts in the range 0.05 $< z <$ 0.3. The lower redshift limit restricted the sample to galaxies with high radio powers (L$_{\rm 1.4GHz} \gtrsim 10^{25}$ W\,Hz$^{-1}$), while the upper redshift limit ensured that the imaging observations had sufficient sensitivity and resolution for the detection of faint interaction signatures. However, the morphologies of 11 of the selected sources have been studied in detail in our previous work \citep{ram11,ram12,ram13}, and were thus excluded from the current sample. One further object could not be used for the classification analysis, since a saturated bright star ruined the image appearance at its location (3C 452). The sample studied here comprises the remaining 72 3CR sources.

\begin{table*}
	\centering
	\caption{\small Basic information for the 72 targets in the 3CR sample described in \S\ref{subsec:3CR_sel}. Full table available as supplementary material.} \label{tab:3CR_info}
	\begin{tabular}{lccccccccc}
	\hline
	Name    & \textit{z} & \makecell{log(L$_{\rm 1.4GHz}$)\\(W\,Hz$^{-1}$)} & \makecell{log(L$_{\rm [OIII]}$)\\(W)} & \makecell{log(M$_{*}$)\\(M$_{\odot}$)} & Optical class  &  Filter & Obs. Date  & \makecell{Seeing \small{FWHM}\\ (arcsec)} \\ \hline 
	3C 20    & 0.174 & 26.88 & 34.52           & 11.3 & HERG & \textit{r} & 2013-08-06 & { 1.28} \\
3C 28    & 0.195 & 26.57 & 33.94            & 11.6 & LERG & \textit{r} & 2012-12-13 & { 2.34} \\
3C 33    & 0.060 &  25.99 & 35.16            & 11.0 & HERG/Q & \textit{r} & 2012-12-13 & { 2.14} \\
3C 33.1  & 0.181 & 26.40 & 35.28            & 11.6 & HERG/Q & \textit{r} & 2013-08-06 & { 1.34} \\
3C 35    & 0.067 & 25.39 & 32.99            & 11.0 & LERG & \textit{r} & 2012-12-13 & { 1.64} \\ 
...    & ... & ... & ...            & ... & ... & ... & ... & ... \\ 
\hline
\end{tabular}
\end{table*}

The spectroscopic properties of the vast majority of the objects in the sample have been determined by \cite{but09,but10,but11}. The measured [OIII]$\lambda$5007 emission-line luminosities or upper limits\footnote{These are 3$\sigma$ upper limits determined from measurements of the noise level in the spectral regions near to the expected positions of the lines, assuming the instrumental resolution as their width \citep{but09}.} from these studies are listed in Table~\ref{tab:3CR_info} (corrected for Galactic reddening), and were only unavailable for three of the sources: the line was not detected in the spectra of 3C 52 and 3C 130; and 3C 405 (Cygnus A) is not included in the Buttiglione et al. samples. For 3C 52 and 3C 130, the upper limits provided for the H$\alpha$ emission-line fluxes were used to estimate [OIII]$\lambda$5007 luminosity upper limits, under the assumption that the noise levels were equivalent in the two spectral regions.

\cite{but09,but10,but11} also used the spectroscopic data to classify the galaxies as HERGs and LERGs, mostly using the excitation index (EI) scheme from \cite{but10}. In the following cases, however, EI indices could not be calculated due to the lack of detections for key spectral lines: 3C 35; 3C 52; 3C 89; 3C 130; 3C 258; 3C 319; 3C 346; and 3C 438. For these objects, we use the [OIII]$\lambda$5007 equivalent width (EW) method adopted by \cite{bh12} to obtain classifications, with HERGs having EW$_{\rm [OIII]} > 5$\AA{} and LERGs having EW$_{\rm [OIII]} < 5$\AA{}. Measurements of the continuum level were estimated from visual inspection of the spectra in \cite{but09} and \cite{but11}. For the objects that lacked clear [OIII]$\lambda$5007 detections, the measured H$\alpha$ fluxes or upper limits were instead used as upper limits on the [OIII]$\lambda$5007 fluxes, for use with the equivalent width criterion (decisive for 3C 52 and 3C 89). When these values were unavailable or were not decisive for the classification, the spectral region near the expected position of the [OIII]$\lambda$5007 line was visually inspected, in order to place an upper limit on its line flux (necessary for 3C 130, 3C 319 and 3C 438).

For 3C 405, the [OIII]$\lambda$5007 emission-line luminosity was calculated from the flux measured by \cite{om75} -- corrected for Galactic reddening using the average $E(B-V)$ value of 0.48 determined by \cite{tay03} from optical spectra of the outer regions of the galaxy -- from which it is identified as a HERG using the EW criterion outlined above. The [OIII]$\lambda$5007 luminosity provided by \cite{but10} for 3C 321 is inconsistent with other measurements in the literature, and we instead adopted the value presented in \cite{dick10}.

\begin{figure}
\centering
    \includegraphics[width=\linewidth]{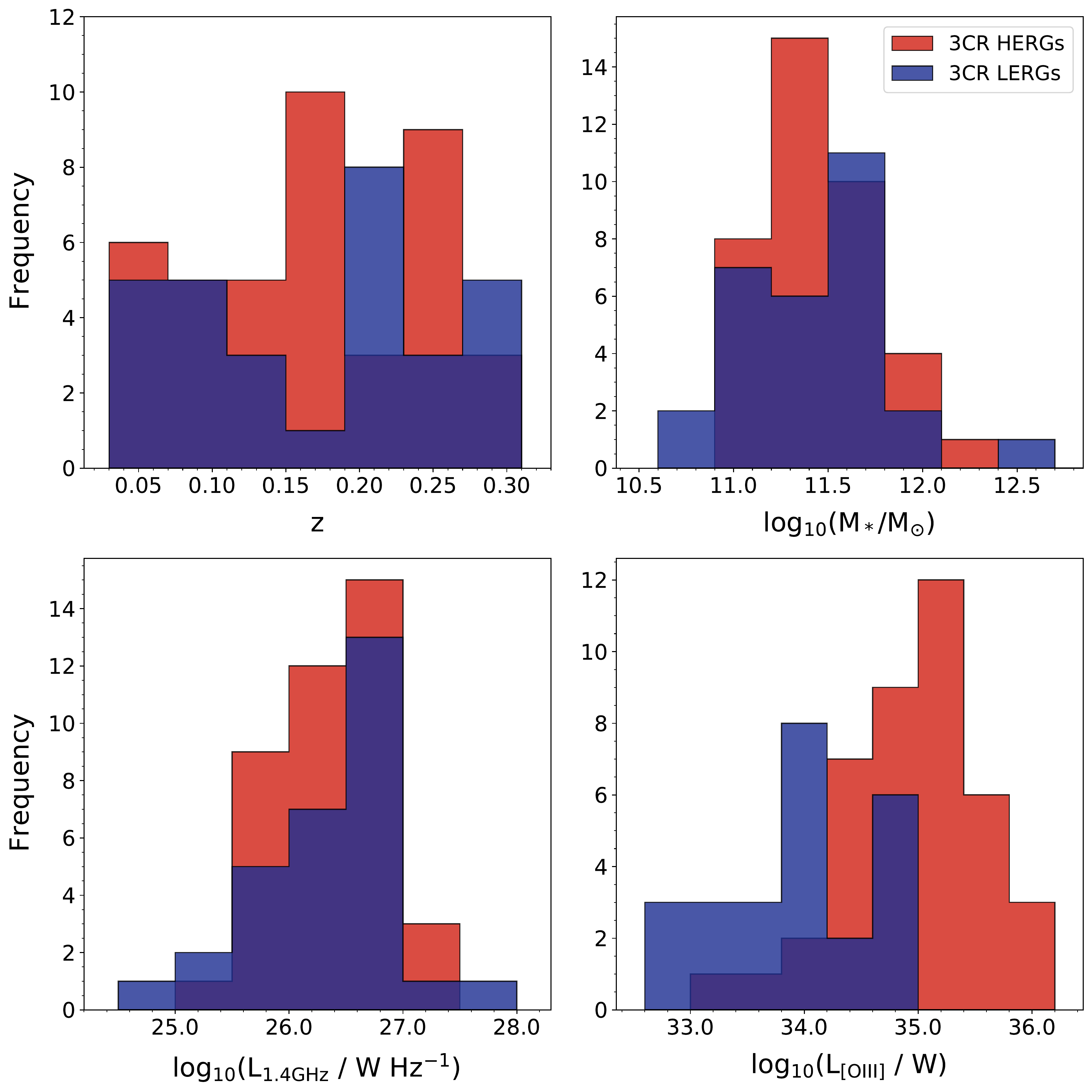}
    \caption{Redshift, stellar mass, 1.4 GHz radio power, and [OIII]$\lambda$5007 emission-line luminosity distributions for the HERG (red) and LERG (blue) sources in the 3CR sample. Objects without mass estimates were not considered when plotting the stellar mass distribution. Those with only upper limits on their [OIII]$\lambda$5007 luminosities were not used for the corresponding distribution.}
   \label{fig:3CR_hists}
\end{figure}

Basic information on the properties of the 3CR sample is presented in Table~\ref{tab:3CR_info}. The 1.4 GHz radio powers were derived from the 178\,MHz flux densities in \cite{spin85}, assuming a standard spectral index of $\alpha=-0.7$. Stellar mass estimates were derived following the procedure outlined in \S\ref{subsec:mass_calc}. Note that we make no distinction between Type 1 and Type 2 AGN in the 3CR sample, since we find that the presence of a broad-line nucleus in the 11 Type 1 objects does not affect the ability to detect large-scale tidal features. The distributions of redshift, stellar mass, 1.4 GHz radio power, and [OIII]$\lambda$5007 emission-line luminosity are shown separately for the 3CR HERGs and LERGs in Figure~\ref{fig:3CR_hists}. 

\subsection{The Type 2 quasar sample}
\label{subsec:q2_sel}

Our Type 2 quasar sample consists of all 25 objects in the SDSS-selected sample of \cite{reyes08} with [OIII]$\lambda$5007 emission-line luminosities above $\rm L_{[OIII]} \geq 10^{8.5}$\,L$_{\odot}$\footnote{Consistent with the Type 2 quasar definition used by \citet{zak03}.} ($\rm L_{[OIII]} \gtrsim 10^{35}$\,W), redshifts $z<0.14$ and right ascension coordinates in the range 00h 30m $< \alpha <$ 12h 30m. These comprise part of a larger sample of 48 Type 2 quasars (the QSOFEED sample), which is described in detail in Ramos Almeida et al. (2021; in preparation).

As for the radio-intermediate HERG samples, the 1.4 GHz radio powers were derived from their NVSS flux densities whenever possible to avoid sensitivity losses due to resolution effects. However, FIRST flux densities were instead used for J0858+31 and J1218+08, where the FIRST images suggested that the NVSS flux densities would be strongly contaminated by other radio sources. FIRST flux densities were also used for J0818+36 and J1015+00, which were not detected by NVSS. J1036+01 and J1223+08 were not detected by either radio survey. For these objects, upper limits on the 1.4 GHz radio powers were estimated using the 1 mJy beam$^{-1}$ flux density selection limit for the FIRST catalogue \citep{white97}. The sample is found to cover a broad range in radio power and includes 21 radio-intermediate sources ($10^{22.5} <$ L$_{\rm 1.4GHz}$ $< 10^{25}$ W\,Hz$^{-1}$) and two traditionally radio-loud sources (L$_{\rm 1.4GHz}$ $> 10^{25}$ W\,Hz$^{-1}$): J1137+61 and 3C 223\footnote{3C 223 was observed and classified twice as part of both the Type 2 quasar sample and the 3CR sample. The object is considered as part of both samples throughout the analysis, although the classifications obtained from the Type 2 quasar image are used when they are combined, due to the higher image quality. The 1.4 GHz radio power and [OIII]$\lambda$5007 luminosity from Table~\ref{tab:3CR_info} are also used for this object in Table~\ref{tab:q2_target_info}, for consistency.}.

\begin{figure}
\centering
    \includegraphics[width=\linewidth]{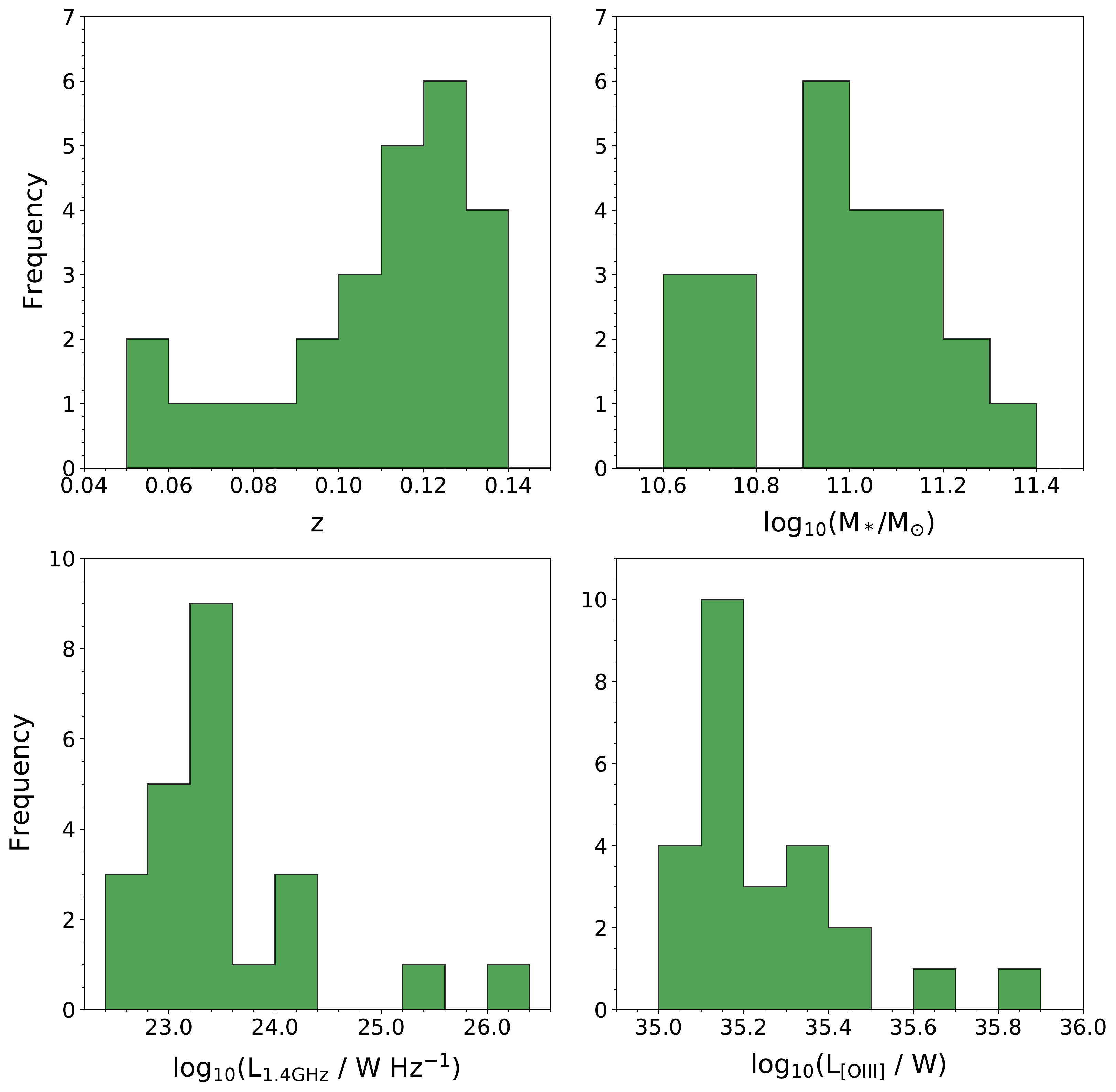}
    \caption{Redshift, stellar mass, 1.4 GHz radio power, and [OIII]$\lambda$5007 emission-line luminosity distributions for the galaxies in the Type 2 quasar sample.}
   \label{fig:q2_hists}
\end{figure}

Calculation of the stellar mass estimates for the hosts of the Type 2 quasars was as performed for the 3CR sources (see \S\ref{subsec:mass_calc}). Basic information for the Type 2 quasar sample is presented in Table~\ref{tab:q2_target_info}. The redshift, stellar mass, 1.4 GHz radio power and [OIII]$\lambda$5007 emission-line luminosity distributions for the targets in the sample are presented in Figure~\ref{fig:q2_hists}.

\begin{table*}
\centering
\caption{\small Basic information for the 25 targets in the Type 2 quasar sample described in \S\ref{subsec:q2_sel}. Full table available as supplementary material.} \label{tab:q2_target_info} 
\begin{tabular}{lcccccccc}
\hline
SDSS ID (Abbr.) & $z$ & \makecell{$A_{\rm r}$\\(mag)} & \makecell{SDSS \textit{r} mag\\(mag)} & \makecell{log(L$_{\rm 1.4GHz}$)\\(W\,Hz$^{-1}$)} & \makecell{log(L$_{\rm [OIII]}$)\\(W)} & \makecell{log(M$_{*}$)\\(M$_{\odot}$)} & Obs. date & \makecell{Seeing \small{FWHM}\\ (arcsec)}\\
\hline 
J005230.59$-$011548.3 (J0052$-$01) & 0.135 & 0.11 & 17.64 & 23.27 & 35.11 & 10.8 & 2020-01-25 & { 1.37} \\
J023224.24$-$081140.2 (J0232$-$08) & 0.100 & 0.09 & 16.65 & 23.04 & 35.13 & 10.8 & 2020-01-25 & { 1.31} \\
J073142.37+392623.7 (J0731+39) & 0.110 & 0.14 & 16.94 & 22.99 & 35.08 & 11.0 & 2020-01-25 & { 1.40} \\
J075940.95+505023.9 (J0759+50) & 0.054 & 0.09 & 15.68 & 23.46 & 35.32 & 10.6 & 2020-01-19 & { 1.47} \\
J080224.34+464300.7 (J0802+46) & 0.121 & 0.17 & 16.94 & 23.48 & 35.11 & 11.1 & 2020-01-25 & { 1.72} \\
... & ... & ... & ... & ... & ... & ... & ... & ... \\ \hline
\end{tabular}
\end{table*}

\subsection{INT/WFC observations}
\label{subsec:obs}

All deep optical imaging data were obtained using the Wide-Field Camera (WFC) attached to the 2.5m Isaac Newton Telescope (INT) at the Observatorio del Roque de los Muchachos, La Palma. The WFC consists of four thinned, anti-reflection-coated 2048 $\times$ 4100 pixel CCDs separated by gaps of $660 - 1098$ \micron, pixel sizes of 0.333 arcsec\,pixel$^{-1}$, and a resulting large total field-of-view of 34 $\times$ 34 arcmin$^{2}$. 
The vast majority of the images were taken using the WFC Sloan \textit{r}-band filter ($\rm{\lambda_{eff}} = 6240$\,\AA{}, $\Delta \lambda = 1347$\,\AA{}), chosen for consistency with the radio-intermediate HERG observations in Paper I and the observations of radio-loud galaxies in the 2 Jy sample, performed by \citet[][]{ram11}. However, images were taken using the WFC Sloan \textit{i}-band filter ($\rm{\lambda_{eff}} = 7743$\,\AA{}, $\Delta \lambda = 1519$\,\AA{}) for six of the 3CR sources (3C 52, 3C 61.1, 3C 63, 3C 130, 3C 405, 3C 410), in order to reduce the influence of the bright moonlight in the period in which these observations were performed. 3C 198 was observed using the WFC Harris R-band filter ($\rm{\lambda_{eff}} = 6380$\,\AA{}, $\Delta \lambda = 1520$\,\AA{}), which has a different response function to the WFC Sloan \textit{r}-band filter but covers a broadly similar wavelength range. 

The observations for the 3CR, RI-HERG high and Type 2 quasar samples were conducted in several separate runs between December 2012 and January 2020, on the dates listed in Tables~\ref{tab:RI-HERG_high_target_info}, \ref{tab:3CR_info} and \ref{tab:q2_target_info}. Observations of the RI-HERG low sample were taken in March 2017, and are described in Paper I. Individual seeing full width at half maximum (FWHM) values for the observations were obtained from averaging the measured FWHM values of foreground stars in the final coadded images (following the reduction outlined in \S\ref{subsec:red}). These values are presented in Tables~\ref{tab:RI-HERG_high_target_info}, \ref{tab:3CR_info} and \ref{tab:q2_target_info}.
The seeing FWHM measurements range from 0.93 to 2.61 arcsec across the four samples, with a median of 1.39 arcsec. The RI-HERG low and RI-HERG high samples were observed with the best typical seeing, with medians of 1.31 and 1.36 arcsec and standard deviations of 0.15 and 0.26 arcsec, respectively. The 3CR and Type 2 quasar samples were typically observed in poorer seeing conditions, with medians of 1.51 and 1.40 arcsec and standard deviations of 0.39 and 0.30 arcsec, respectively. A summary of the seeing measurements for each sample is presented in Table~\ref{tab:lsb_summary}.

The vast majority of the targets were observed using 4 $\times$ 700s exposures, yielding total exposure times of 2800s per target. These integration times were necessary for the detection of low-surface-brightness tidal features, but were divided into separate exposures to avoid target galaxy saturation in individual images. A square dithering pattern was employed for the observations (four pointing offsets of 30 arcsec) in order to cover the gaps in the images introduced by the spacings between the CCDs. Flat-field and image defect corrections were also improved by this process. 

Several targets were, however, observed with different total integration times. 3C 405 and 3C 410 lie in crowded fields that contain a large number of stars due to their close proximity to the Galactic plane. The observations of these targets were divided into additional separate exposures that would yield similar total exposure times and reduce the number of saturated stars in individual images: 5 $\times$ 500s for 3C 405 and 7 $\times$ 400s for 3C 410. Only three exposures were obtained for 3C 63 (3 $\times$ 700s), J0818+36 (3 $\times$ 700s) and 3C 458 (3 $\times$ 600s), while five were used for 3C 236, 3C 326 and J0945+17 (each 5 $\times$ 700s), six for 3C 349 (6 $\times$ 700s) and seven for J1015+00 (7 $\times$ 700s). Four shorter exposures were taken for J1223+08 (4 $\times$ 400s), due to time constraints on the night on which this target was observed. In addition, image quality issues meant that only 3 $\times$ 700s individual frames could be used for some targets: J0836+05 and 3C 236 (satellite trails); J0858+31, J0945+17 and J1200+31 (bad CCD striping). Targets with three observations were still dithered sufficiently to overcome the issues related to the WFC CCD gaps.  

\subsection{Image reduction and surface brightness depth}
\label{subsec:red}

All processing of the target images, from initial reduction to construction of the final mosaic images, was carried out using \texttt{THELI} (\citeauthor{schirmer13}, \citeyear{schirmer13}; latest version available at \url{https://github.com/schirmermischa/THELI}). 
Bias and flat-field corrections for each observation were performed via the subtraction of master bias frames and division by the master flat-field frames, respectively.
For the \textit{i}-band observations, fringing effects were removed using a two-pass image background model. Brighter background features were removed on the first pass, where no detection and masking of objects was performed. Fainter background features were modelled in the second pass, performed after the detection and masking of all objects in the field above a certain threshold (signal-to-noise per pixel of 1.5 for objects with 5 or more connected pixels).

Astrometric solutions for the 3CR, RI-HERG high and Type 2 quasar calibrated images were calculated in \texttt{THELI} through cross-matching object catalogues produced for each image by \texttt{SExtractor} \citep[][]{bertin96} with the GAIA DR2 catalogue (\citeauthor{gaia18}, \citeyear{gaia18}). The all-sky USNO-B1 catalogue (\citeauthor{monet03}, \citeyear{monet03}) was used for the RI-HERG low sample, since these data were reduced with an earlier version of \texttt{THELI} (see Paper I).
A sky model was produced and subtracted from all calibrated images by \texttt{THELI} to correct for remaining variations in the sky background level before final coaddition.

Photometric zero points were calculated in \texttt{THELI} through comparison of the derived instrumental magnitudes for stars in the final coadded image fields with their catalogued magnitudes in the Pan-STARRS1 \citep{cham16} First Data Release\footnote{Note that the analysis in Paper I considered catalogued SDSS \textit{r}-band magnitudes when determining the zero points for the RI-HERG low sample. However, the Pan-STARRS1 magnitudes are used when comparing between the different samples in the current work. Both surveys use the AB magnitude system and the difference between the \textit{r}-band magnitudes is minor, characterised by the following equation: $r_{\rm SDSS} = r_{\rm P1} - 0.001 + 0.011(g-r)_{\rm P1}$ \citep{tonry12}.}. This method has the key advantage of automatically correcting for photometric variability throughout the nights, since the calibration stars are observed at the same time and at the same position on the sky as the galaxy targets. The reliance on average zero points derived from standard star observations is also removed.

Measurements of the surface brightness depth of the observations were obtained using the same procedure as outlined in Paper I, which closely follows that used by \citet{atk13}. These measurements were carried out after all reduction had been performed, including the flattening achieved through the subtraction of the sky background model. The total counts detected within 40 unique circular apertures of one arcsecond in radius were measured, placed in regions of the sky background with minimal or no influence from the haloes of bright objects in the field.

The standard deviations in the sky background level within the apertures ($\sigma_{\rm sky}$) were converted to apparent magnitudes, using the photometric zero points derived as outlined above, to provide final measurements of the surface brightness depth. The median $3\sigma_{\rm sky}$ surface brightness depths over all \textit{r}-band observations of the targets is 27.0 mag\,arcsec$^{-2}$, with a standard deviation of 0.3 mag\,arcsec$^{-2}$. The value for the \textit{i}-band observations of the 3CR targets is 25.7 mag\,arcsec$^{-2}$ ($3\sigma_{\rm sky}$), with a standard deviation of 0.2 mag\,arcsec$^{-2}$. The differences in individual and total exposure times for some target observations (see \S\ref{subsec:obs}) did not significantly affect the surface brightness depths achieved. Limiting surface brightness measurements for individual targets in the RI-HERG low sample were presented in Paper I. Values for the individual target observations are not presented for the RI-HERG high, 3CR and Type 2 quasar samples. However, a summary of the average values for each sample is presented in Table~\ref{tab:lsb_summary}.

\begin{table}
\small
\centering
\caption{\small The means, medians and standard deviations of the seeing FWHM and $3\sigma_{\rm sky}$ limiting surface brightness measurements ($\mu_{AB}$) for the observations of the four samples. The latter values have units of mag\,arcsec$^{-2}$ and are presented in the Pan-STARRS1 AB magnitude photometric system.}
\label{tab:lsb_summary}
\begin{tabular}{ccccccc}
\hline
                                  &      & Seeing &          &      & $\mu_{AB}^{3\sigma_{\rm sky}}$ &          \\
                                  & Mean & Median                        & $\sigma$ & Mean & Median                         & $\sigma$ \\ \hline
\makecell{RI-HERG low}           & 1.42 & { 1.36}                          & 0.26      & 27.0 & 27.0                           & 0.3      \\
\makecell{RI-HERG high}          & {1.25} & { 1.31}                          & {0.15}      & 27.1 & 27.1                           & 0.3      \\
\makecell{Type 2 quasars}        & { 1.44} & { 1.40}                         & { 0.30}      & 27.0 & 27.0                           & 0.3      \\
\makecell{3CR (\textit{r}-band)} & { 1.67} & { 1.51}                          & { 0.40}      & 26.9 & 27.0                           & 0.4      \\
\makecell{3CR (\textit{i}-band)} & { 1.42} & {1.47}                          & 0.12      & 25.7 & 25.7                           & 0.2  \\   \hline
\end{tabular}
\end{table}

\section{Methodology and control matching}
\label{sec:methods}

\subsection{Stellar mass calculation}
\label{subsec:mass_calc}

Prior to identifying suitable matched control galaxies, estimates of the stellar masses were required for all of the targets to be matched. All objects in the RI-HERG low and RI-HERG high samples had existing stellar mass estimates in the MPA-JHU value added catalogue from which the control galaxies were also extracted\footnote{These were obtained from fitting of the stellar population synthesis models of \cite{bc03} to the SDSS broad-band \textit{ugriz} photometry for the galaxies. The method is similar to that described by \cite{kauf03}, who instead used spectroscopic features to characterise the fits. A comparative discussion of these methods is available at: \url{https://wwwmpa.mpa-garching.mpg.de/SDSS/DR7/mass_comp.html}.}, and so no additional calculations were required for these sources. 
However, MPA-JHU measurements were not available for all of the objects in the 3CR and Type 2 quasar samples. Initial stellar mass estimates for the galaxies in these samples were instead derived from their Two Micron All Sky Survey \citep[2MASS;][]{skrut06} $K_s$-band luminosities, using the colour-dependent mass-to-light ratio prescription of \cite{bell03} for this waveband. This method was used to derive stellar mass estimates for \textit{all} of the objects in both the 3CR and Type 2 quasar samples (i.e. MPA-JHU values were not used for any of these galaxies), for consistency both within and between the two samples.

The \cite{bell03} mass-to-light ratio prescription for the 2MASS $K_s$-band has the following form (in solar units):
\vspace{-0.2ex}
\begin{equation}
    \label{eq:m-to-L}
    \mathrm{log}(M/L)_{K_s} = -0.206 + 0.135 \times (B-V)\,.
\end{equation}

\noindent
A $B-V$ colour of 0.95 was assumed for all sources, in line with the expected value for a typical elliptical galaxy at zero redshift \citep[e.g.][]{sh89b}. An additional factor of 0.15 dex was subtracted in order to convert to a Kroupa initial mass function (IMF) for the stellar population \citep{kroupa01}, in accordance with the IMF assumed for the MPA-JHU stellar mass estimates. A final value of $\mathrm{log}(M/L)_{K_s} = -0.228$ was thus used for the calculations.

Where possible, the ``total" $K_s$-band magnitudes\footnote{Derived by extrapolating the radial surface brightness profile measured within a 20 mag\,arcsec$^{-2}$ elliptical isophote to a radius of around four scale lengths (from S\'ersic-like exponential fitting), in an attempt to account for the flux lost below the background noise.} listed in the 2MASS Extended Source Catalog \citep{jarr00} were used to derive the luminosities for the calculations, in order to ensure that as much as possible of the galaxy flux was encapsulated (XSC magnitudes, hereafter). These values were available for 44 of the 73 3CR sources (60 per cent) and 19 of the 25 Type 2 quasars (76 per cent). For the remaining sources in the two samples, the magnitudes listed in the 2MASS Point Source Catalog \citep[][]{skrut06}, derived from PSF profile fitting, were considered (hereafter PSC magnitudes). These were available for all 6 of the remaining Type 2 quasars and 26 of the 29 remaining 3CR galaxies (90 per cent). 

Three of the 3CR sources -- 3C 61.1, 3C 258, and 3C 458 -- did not have either 2MASS XSC or PSC magnitudes available. Stellar mass estimates were not calculated in these cases, and, consequently, the control matching was not performed for these sources. However, these objects were still considered for the morphological classifications and are included in comparisons of the results obtained for the 3CR sample and their matched controls (\S\ref{sec:res}). This was done under the assumption that the stellar masses for these sources, which represent a small minority of the full 3CR sample (4 per cent), are similar to those of the other 3CR objects.

\begin{figure}
    \centering
    \includegraphics[width=\linewidth]{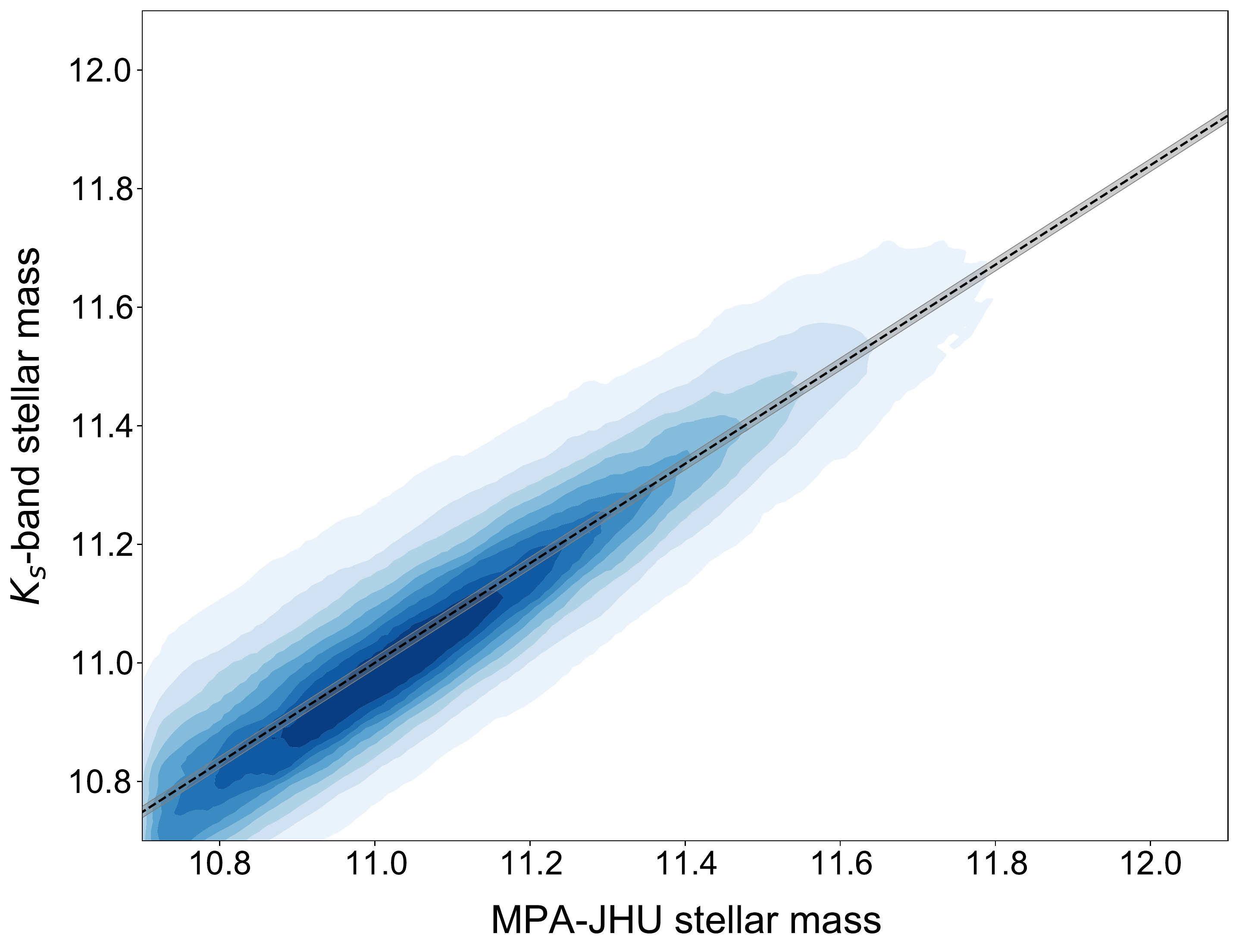}
    \caption[]{A plot of the $K_s$-band and MPA-JHU stellar mass estimates for the 238,418 galaxies in the MPA-JHU catalogue with 2MASS XSC magnitudes and MPA-JHU stellar mass estimates in the range 10.7 $\leq$ $ \rm log(M_*/M_{\odot}$) $\leq$ 12.0. The line of best fit derived from linear regression is plotted (black), with the grey shaded region representing the one-$\sigma$ error bounds for the fit.}
    \label{fig:k_mass_v_mpa_mass}
\end{figure}

The main disadvantage of using the PSC measurements was loss of sensitivity to emission from the extended regions of the target galaxies. In an attempt to account for this effect, a correction was derived from the average difference between the XSC and PSC magnitudes for the large number of galaxies in the MPA-JHU catalogue that had both measurements available. Considering all MPA-JHU galaxies with redshifts in the range covered by both the 3CR and Type 2 quasar samples ($0.05 < z < 0.3$), a median magnitude difference of $K_s^{\rm PSC} - K_s^{\rm XSC} =0.811$ was determined, with a standard deviation of 0.311. 
No significant evidence for a relationship between these differences and the measured PSC magnitude or redshift was found, as confirmed by Pearson correlation tests ($r = -0.121$ and $r = 0.100$, respectively, with negligible \textit{p}-values). As a result, a fixed value of 0.811 was subtracted from the PSC magnitudes to convert them to estimated XSC magnitudes in all cases, with the latter then being used in the subsequent calculations.

Following this, an extragalactic extinction correction was subtracted from the XSC or corrected PSC magnitudes, using the \textit{K}-band values of \cite{sf11}\footnote{Downloaded from the IRSA Galactic Dust Reddening and Extinction online service, available at: \url{https://irsa.ipac.caltech.edu/applications/DUST/.}}. A cosmological \textit{k}-correction was also applied, using the $k(z) \approx (2.1 \pm 0.3)z$ formulation given in \cite{bell03}, which is independent of galaxy spectral type. The corrected magnitudes were then converted to solar luminosities by assuming a solar 2MASS \textit{K$_s$}-band absolute magnitude of 3.27  \citep[][]{will18}, after which Equation~\ref{eq:m-to-L} was applied to obtain the stellar mass estimates. An AGN contribution to the near-infrared light was not subtracted before calculating the stellar masses. However, we do not expect this to have a strong influence on our main results, since: (i) high-resolution near-IR imaging of nearby 3CR radio galaxies showed that the contribution from Type 2 AGN was typically small near $K_s$-band wavelengths \citep[$<$20 per cent of the light in 0.9” diameter nuclear apertures;][]{ram14a,ram14b}; (ii) Type 1 AGN objects only comprise 15 per cent of the 3CR sample (11/72) and 7 per cent of the full AGN sample (11/154), and Type 1 quasars only 7 per cent and 3 per cent, respectively.

As a final step, the same technique was used to determine $K_s$-band stellar mass estimates for all galaxies in the MPA-JHU catalogue with 2MASS XSC magnitudes available. This was done with the goal of obtaining a correction to align the $K_s$-band stellar mass estimates with the MPA-JHU values. Since individual SDSS $g-i$ colours were available for the MPA-JHU galaxies, a second prescription from \cite{bell03} that utilized these measurements was employed in this instance: $\mathrm{log}(M/L)_{K_s} = -0.211 + 0.137 \times (g-i)$. The procedure was otherwise identical to that described above; this again included the subtraction of an additional 0.15 dex to convert to a Kroupa IMF. All XSC-matched galaxies with MPA-JHU stellar mass estimates in the range 10.7 $\leq$ $ \rm log(M_*/M_{\odot}$) $\leq$ 12.0 were considered for the comparison, in approximate agreement with the range of $K_s$-band stellar mass estimates derived for the 3CR and Type 2 quasar samples -- while some of the 3CR sources had $K_s$-band estimates larger than $ \rm log(M_*/M_{\odot}$) = 12.0, this upper limit was chosen because of the large uncertainties determined for MPA-JHU estimates above this value. 

The $K_s$-band and MPA-JHU stellar mass estimates for the $\sim$240,000 galaxies considered for the comparison are presented in Figure~\ref{fig:k_mass_v_mpa_mass}. The line of best fit displayed in the figure has the form $ \rm log(M_*/M_{\odot})_{K_s}$ = 0.84\,$ \rm log(M_*/M_{\odot})_{\rm MPA-JHU}$ + 1.77, as derived from linear regression. The typical scatter around the relation is 0.01. No significant evidence that the difference between the stellar mass estimates varies with $g-i$ colour, redshift, 2MASS XSC $K_s$-band magnitude, or the values of either of the stellar mass estimates was found, based on Pearson correlation tests. Consequently, this relation was employed to convert $K_s$-band stellar mass estimates to corresponding MPA-JHU stellar mass estimates in all cases. The final MPA-JHU-equivalent stellar mass estimates for the 3CR and Type 2 quasars are presented in Tables~\ref{tab:3CR_info} and \ref{tab:q2_target_info}, respectively.

\subsection{Control matching procedure}
\label{subsec:control_matching}

The MPA-JHU value-added catalogue contains a large amount of raw and derived data for 927,552 galaxy spectra from SDSS DR7, including the spectroscopic redshifts and stellar mass estimates that were crucial for the control matching. Given the significant crossover between the fields covered by the INT/WFC images and the SDSS DR7 survey footprint, the catalogue therefore provided a suitable ``pool" from which to select matched control galaxies for the active galaxies with matching observation properties (technique, image depth, observing conditions). Prior to performing the control matching, however, the following steps were taken to limit the control pool to galaxies with suitable properties.

\begin{enumerate}
    \item \textit{Coordinate restriction.} Galaxies were required to lie within the regions of sky covered by the INT/WFC images of the active galaxies. This was the most restrictive constraint, with only 2,744 of the objects in the catalogue meeting this criterion (0.3 per cent).
    \item \textit{Removal of likely AGN.} The MPA-JHU catalogue was constructed solely of objects that had been spectroscopically classified as galaxies in SDSS DR7, with the exception of some objects that had originally been targeted as galaxies but were later classified as quasars. However, the spectral types of both galaxies and quasars were also subclassified based on the properties of their [OIII]$\lambda$5007, H$\alpha$, H$\beta$, and [NII]$\lambda$6583 emission lines (if strongly detected).
    All sources that had been subclassified as either AGN or non-star-forming broad-line objects\footnote{Objects with lines detected at the 10\,$\sigma$ level, with velocity dispersion $> 200$ km\,s$^{-1}$ at the 5\,$\sigma$ level.} were removed from consideration. Star-forming or starburst galaxies were not removed. This left 2,615 galaxies in the control pool.
    \item \textit{Removal of remaining targets.} The three galaxies from our target samples that had not been identified as AGN based on their SDSS spectra in step (ii) were removed, leaving 2,612 objects.
    \item \textit{Removal of duplicate objects.} The MPA-JHU catalogue contains duplicate identifications for a large number of galaxies due to repeat SDSS spectral observations. Any remaining duplicates were taken out of the control pool at this point, which left 2,413 sources available for the matching.
\end{enumerate}

All matched controls used for the analysis were selected from this final restricted pool of 2,413 galaxies. As mentioned, this matching was not performed for the three galaxies in the 3CR sample for which stellar mass estimates could not be obtained from 2MASS magnitudes (\S\ref{subsec:mass_calc}): 3C 61.1; 3C 258; and 3C 458. The Type 2 quasars were also not considered at this stage, although matching was performed for these objects after the classifications had been obtained (see below). A total of 127 active galaxies (58 radio-intermediate HERGs and 69 3CR sources) were therefore considered for control matching. Repeat selections of controls that matched multiple active galaxies were permitted throughout the matching process, in order to maximise the number of possible matches available for each target.

Redshift matches were determined using separate criteria for the radio-intermediate HERG and 3CR samples:

\begin{enumerate}
    \item $z_{\rm target} - 0.01 < z_{\rm control} < z_{\rm target} + 0.01$, for matching to the RI-HERG low and RI-HERG high samples;
    \item $z_{\rm target} - 0.02 < z_{\rm control} < z_{\rm target} + 0.02$, for matching to the 3CR sample.
\end{enumerate}

\noindent
An increased tolerance was allowed for the 3CR matching due to the fact that their typically high stellar masses (median log(M$_*$/M$_{\odot}$) = 11.4) made the selection of matched controls more difficult (see below). All control galaxies with suitable redshifts then needed to meet both of the following two stellar mass criteria:

\begin{enumerate}
    \item (log(M$_*$/M$_{\odot}$) + $\sigma$)$_{\rm control}$ $>$ (log(M$_*$/M$_{\odot}$) $-$ $\sigma$)$_{\rm target}$\,;
    \item (log(M$_*$/M$_{\odot}$) $-$ $\sigma$)$_{\rm control}$ $<$ (log(M$_*$/M$_{\odot}$) $+$ $\sigma$)$_{\rm target}$\,;
\end{enumerate}

\noindent
i.e. the $1\sigma$ uncertainties on the stellar mass estimates for the target and the control were required to overlap \citep[as in][]{gord19}. A total of 1,581 unique control galaxies were found to meet these selection criteria, an average of $\sim$12 per active galaxy considered. However, since repeat selections of controls were permitted, the true number of matches was in fact much larger than this (8,700).

In addition to the three 3CR sources not considered for the matching (listed above), no matches were found for a further 10 of the 3CR sources using these criteria: 3C 130; 3C 234; 3C 323.1; 3C 346; 3C 371; 3C 382; 3C 405; 3C 410; 3C 430; and 3C 433. These sources had stellar mass estimates in the range $\rm 11.3 \leq log(M_*/M_{\odot}) \leq 12.7$, with a median of 11.8, and the lack of success in the matching was found to be driven by the large stellar masses of these objects. Overall, the procedure was thus successful in finding controls for 117 out of the 127 targets considered for the matching (92 per cent).

\begin{figure}
    \centering
    \includegraphics[width=\linewidth]{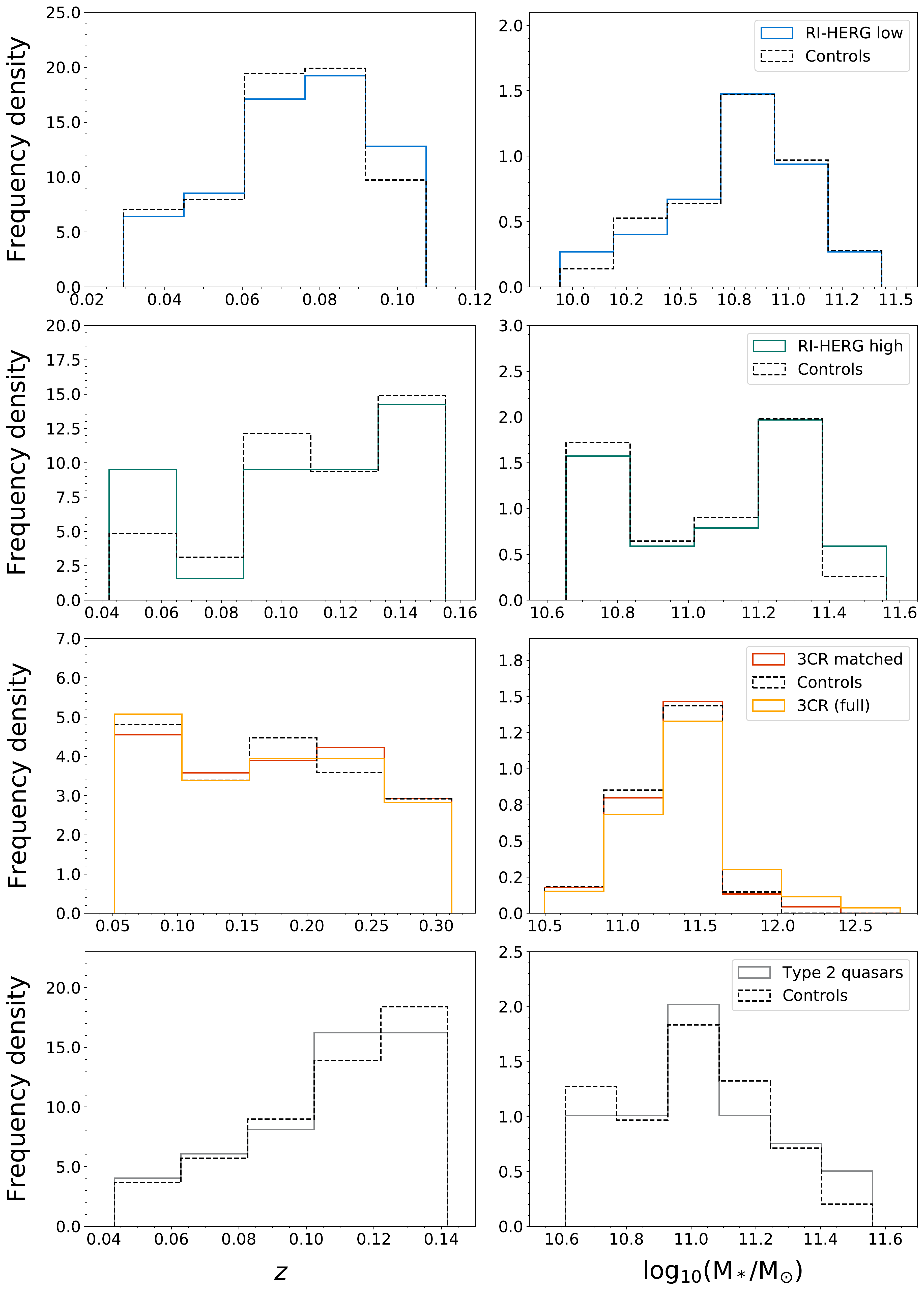}
    \caption{The redshift (left column) and stellar mass (right column) distributions for the galaxies in each of the four active galaxy samples, alongside the corresponding distributions for their matched controls. For the 3CR sample, { the full sample and} the objects successfully matched with controls are { shown separately}.}
    \label{fig:control_matching_hists}
\end{figure}

\begin{table}
\centering
\caption{The results of two-sample Kolmogorov-Smirnoff (KS) tests performed for the redshift and stellar mass distributions of the active galaxy samples and their respective matched control samples. Both the test statistic ($D$) and the $p$-value are presented in each case. Note that the values listed for the 3CR sample we derived considering only the objects for which control matches were successfully found.} 
\label{tab:control_matching_KS}
\begin{tabular}{ccccc}
\hline 
                  & \multicolumn{2}{c}{$z$} & \multicolumn{2}{c}{$ \rm log(M_*/M_{\odot}$)} \\ 
                  &  $D$  &  $p$          & $D$  &  $p$            \\  \hline     
RI-HERG low     &  0.084  & 0.988  & 0.076 & 0.996 \\
RI-HERG high    &  0.119 &  0.857  & 0.163 & 0.519 \\
3CR (matched)   &  0.075 & 0.922  &  0.095 & 0.726 \\
Type 2 quasars  &  0.099 & 0.971  &  0.115 & 0.913 \\
 \hline
\end{tabular}
\end{table}

In order to reduce the total number of galaxies requiring classifications, only the five controls with the smallest differences between the target and control stellar mass estimates were considered for each active galaxy. This requirement had the additional advantage of counteracting the potential selection of controls with large uncertainties on their stellar mass estimates. In cases where the target had fewer than five matches, all of the available controls were considered: J1036+38 (RI-HERG low), with 4 matches; 3C 52, with 3 matches; 3C 236, with 2 matches; and 3C 132, 3C 388, 3C 438, J1630+12 (RI-HERG low), J0752+45, J1147+35, and J1436+05 (RI-HERG high), with one match each. A total of 551 control selections (388 unique galaxies, 163 repeats) were made for the 118 matched targets remaining at this point, an average of 4.7 per active galaxy.

19 of these control galaxies were found to be unsuitable for the classifications due to issues with the images (e.g. bad image regions, defects, crowding/source confusion). In these cases, images for the next closest matches in stellar mass were inspected until a suitable replacement control was found. All 19 controls were successfully replaced, but with 10 additional repeat selections (551 total selections with 173 repeats). This left a final sample of 378 unique control galaxies with images to be used for the classification analysis. 

Due to their late inclusion in the project, control matching for the Type 2 quasar sample was performed after the interface classifications had been obtained. The controls for these targets were hence selected from the 378 control galaxies selected as matches to objects in the other active galaxy samples. The criteria used for the matching were identical to those used for the RI-HERG low and RI-HERG high samples: (i) $z_{\rm target} - 0.01 < z_{\rm control} < z_{\rm target} + 0.01$, for the redshift matching; (ii) (log(M$_*$/M$_{\odot}$) + $\sigma$)$_{\rm control}$ $>$ (log(M$_*$/M$_{\odot}$) $-$ $\sigma$)$_{\rm target}$ and (log(M$_*$/M$_{\odot}$) $-$ $\sigma$)$_{\rm control}$ $<$ (log(M$_*$/M$_{\odot}$) $+$ $\sigma$)$_{\rm target}$, for the stellar mass matching. 

Using these criteria, it was found that 202 of the 378 controls were also matches to at least one of the galaxies in the Type 2 quasar sample. This included matches for each of the Type 2 quasar objects, with the number of matches for each individual target ranging from 4 to 47. As before, the five controls with the smallest differences between active galaxy and control galaxy stellar mass were selected for the control sample, and repeat selections were allowed. Only four matched controls were available for J1100+08, which were hence all included in the sample. A total of 124 control selections were therefore made for the 25 Type 2 quasar objects (91 unique control galaxies, 33 repeats).

Figure~\ref{fig:control_matching_hists} shows the stellar mass and redshift distributions for each of the active galaxy samples and their respective matched control galaxy samples, which demonstrate the success of the matching.
Two-sample Kolmogorov-Smirnov tests performed on these distributions provided no significant evidence for rejecting the null hypothesis that the targets and their matched controls are drawn from the same underlying distribution, in any of the cases. The results of these tests are presented in Table~\ref{tab:control_matching_KS}.

\subsection{Online classification interface}
\label{subsec:interface}

The online interface used to obtain morphological classifications for the project was made using the Zooniverse Project Builder platform at Zooniverse.org\footnote{Available at: \url{https://www.zooniverse.org/lab}.}, a citizen science web portal that stemmed from the initial Galaxy Zoo project by \cite{lin08,lin11}. Through this interface, eight researchers (all authors except PB, PDG) were blindly shown images of the active galaxies and control galaxies in a randomised order, and were asked to answer multiple choice questions concerning their optical morphologies.

No additional information on the nature of the galaxy to be classified was provided, in order to avoid introducing any biases related to the individual galaxy properties (e.g. active/non-active, target name, stellar mass, redshift, radio power, optical luminosity). However, two scale bars of 10 kpc in size were included on each image to assist with determination of multiple nuclei classifications (see below). The images were centred on the targets and fixed to be of 200 kpc $\times$ 200 kpc in size at the redshift of the galaxy in question, in all cases \citep[as in][]{gord19}. The images used for the classifications for all active galaxies (except those in the RI-HERG low sample) are available in the supplementary information.

One disadvantage of carrying out the classifications in this way was that the ability to fully manipulate the image contrast levels \citep[as for those in][]{ram11,ram12,bess12,pierce19,ell19} was lost. The interface was set up with several features that partially accounted for this issue. Firstly, two postage stamp images with different contrast levels were displayed alongside each other for each galaxy: one of high contrast, for clearer identification of high-surface-brightness tidal features and the overall morphological types (spiral/disk, elliptical, etc.); and one of low contrast, for clearer identification of faint morphological structures. The two image contrast levels were chosen manually on a case-by-case basis, with consideration given to the appearance of both the target galaxy structures and the objects and/or image defects within the 200 kpc square surrounding region. In addition, classifiers could zoom in or out and pan around the image to look at specific regions in more detail. Rotation or inversion of the image was also possible.

Using the interface, classifiers were required to answer up to three multiple choice questions related to the morphological appearance of the subject galaxies. The first of these required the classifiers to answer the question ``Does this galaxy show at least one clear interaction signature?" using one of the following options: (i) ``Yes"; (ii) ``No"; or (iii) ``Not classifiable (e.g. image defect, bad region/spike from saturated star)". 
The last option was included for cases in which it was not possible to determine whether or not the galaxy was disturbed due to issues with the displayed image (e.g. because of presence of major image defects). For consistency with the previous study of the RI-HERG low sample (Paper I), dust lanes were included as one of the interaction signature classifications at this stage. However, dust lanes were not considered as a clear signature of a galaxy merger or interaction, and these classifications were not used for the analysis presented in \S\ref{subsec:dist_rates} and \S\ref{subsec:features_merger_stage}.

Should the classifier have answered ``Yes" to this first question, they were then asked to identify the type(s) of interaction signature that they had seen. To do this, they had to answer the question ``What types of interaction signature are visible?" using the following options:

\begin{enumerate}
    \item Tail (T) -- a narrow curvilinear feature with roughly radial orientation;
    \item Fan (F) -- a structure similar to a tail, but that is shorter and broader;
    \item Shell (S) -- a curving filamentary structure with a roughly tangential orientation relative to a radial vector from the main body of the galaxy;
    \item Bridge (B) -- a feature that connects a radio galaxy with a companion;
    \item Amorphous halo (A) -- the galaxy halo is misshapen in an unusual way in the image;
    \item Irregular (I) -- the galaxy is clearly disturbed, but not in a way that fits any of the alternative classifications;
    \item Multiple nuclei (2N, 3N...) -- two or more brightness peaks within a distance of 10\,kpc;
    \item Dust lane (D) -- a clear linear dark structure within the galaxy;
    \item Tidally interacting companion (TIC) -- a companion galaxy shows clear morphological disturbance that is suggestive of a tidal interaction with the main target (e.g. with direction aligned towards/away from the central target).
\end{enumerate}

\noindent 
The classifiers were allowed to select as many options as necessary, to ensure that multiple interaction signatures could be identified for each galaxy when present. These categories were chosen to be consistent with those from the interaction signature classification scheme detailed in Paper I \citep[also with][]{ram11,ram12,bess12}. 
A new category was also added to the classification scheme for the current analysis: ``Tidally interacting companion (TIC)". This accounted for cases in which a close companion showed evidence for a tidal interaction with the main target, whether or not the target itself showed clear interaction signatures -- this included cases where the distance limit criterion for the ``Multiple Nuclei (2N, 3N...)" class was not met. 

Finally, the classifiers were required to answer the question ``On first impression, what is the morphological type of the galaxy?" using one of the following responses: (i) ``Spiral/disk"; (ii) ``Elliptical"; (iii) ``Lenticular"; (iv) ``Merger (too disturbed to classify as above)"; or (v) ``Unclassifiable (due to image defects, \textit{not} merger)".
Again, these options were chosen to be consistent with the host type classifications obtained from visual inspection of the RI-HERG low sources in Paper I. As for the first question, the last option was included for cases where the classifier thought that issues with the image quality meant that they could not provide an accurate classification.

\section{Analysis and results}
\label{sec:res}

Through the online interface, eight researchers (all authors except PB, PDG) provided morphological classifications for each of the 533 galaxies involved in the project. In the same manner as the morphological classification analysis performed for the RI-HERG low sample (Paper I), each classification was considered as a ``vote" for that particular classification category. A classification was then only accepted when the number of votes it received exceeded a certain threshold, the value of which was dependent on the question considered. In this section, the results related to each of the three classification questions listed in \S\ref{subsec:interface} are addressed in turn. Full classification results for each of the 155 active galaxies studied are presented in the supplementary information.

\subsection{Rates of morphological disturbance}
\label{subsec:dist_rates}

The main goal of the project was to determine how the importance of galaxy mergers and interactions for triggering AGN varies with their radio powers and/or optical emission-line luminosities. This topic was addressed by the first question asked to the classifiers in the online interface: ``Does this galaxy show at least one clear interaction signature?", with the possible answers of ``Yes", ``No" and ``Not classifiable (e.g. image defect, bad region/spike from saturated star)" (see \S\ref{subsec:interface}).

This question was answered by all eight classifiers for every galaxy in the sample. Only one of the listed responses could be selected when answering this question. A threshold of 5 out of 8 votes was chosen as the lower limit for accepting a certain classification in all cases (i.e. a simple majority). In this instance, the goal was simply to test whether or not the classifier believed that the galaxy had been disturbed by a merger or interaction. Here, cases in which 5 or more votes were recorded for ``Yes" were taken to confirm that the galaxy was disturbed, and those where 5 or more votes were recorded for ``No" were taken to indicate that the galaxy was not disturbed. Any other distribution of votes (including any number for ``Not classifiable") was considered as an uncertain case.

\subsubsection{Proportions}
\label{subsubsec:q1-props}

\begin{figure}
    \centering
	\includegraphics[width=\columnwidth]{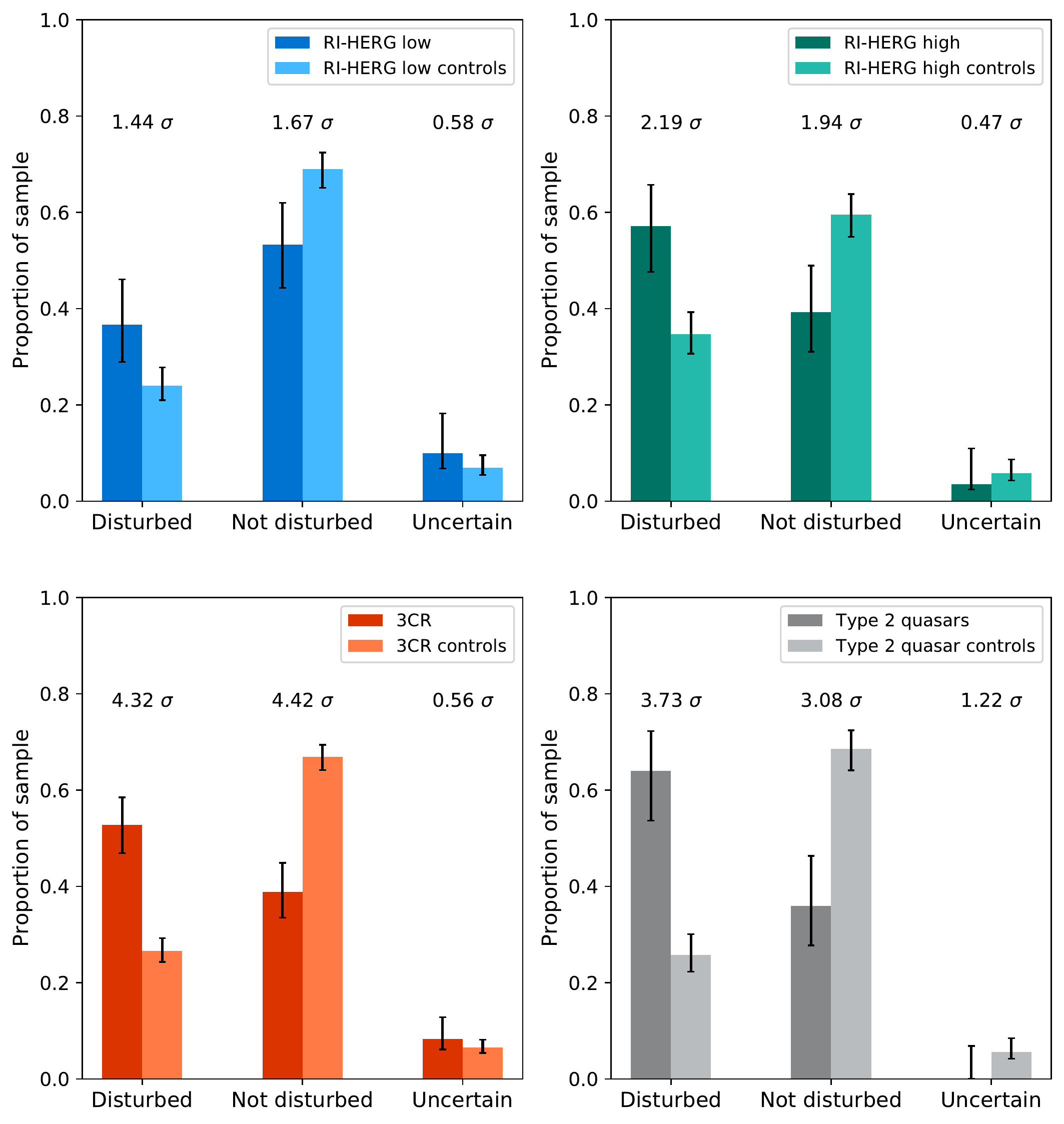}
    \caption{The proportions of galaxies classed as disturbed, not disturbed or uncertain for each of the samples classified using the online interface. The results for the active galaxy samples are presented alongside those for their respective matched control samples, with the significance of the difference between the measured proportions for each category indicated in all cases (from two-proportion Z-tests). Proportion uncertainties were estimated following the method of \citet[][]{cam11}. The exact proportions are presented in Table~\ref{tab:q1_props}.}
    \label{fig:q1_all_agn_samples}
\end{figure}

Figure~\ref{fig:q1_all_agn_samples} shows the proportions of galaxies classed as disturbed, not disturbed and uncertain for all samples classified using the online interface. The results for the active galaxy samples are presented alongside those for their matched control samples in all cases. The significance of the differences between the active galaxy samples and matched control samples is also shown, as estimated using the two-proportion Z-test. The proportions measured for each of the active galaxy and matched control samples are presented in Table~\ref{tab:q1_props}.

\begin{table}
\centering
\caption{The proportions of galaxies classed as disturbed, not disturbed and uncertain for all of the active galaxy and matched control samples classified using the online interface, as presented in Figure~\ref{fig:q1_all_agn_samples}. All proportions are expressed as percentages. The results for the 3CR HERG and LERG subsamples are included in separate columns, alongside those found for the full 3CR sample. Proportion uncertainties were estimated following the method of \citet[][]{cam11}.} 
\label{tab:q1_props}
\begin{tabular}{C{1.13cm}C{0.88cm}C{0.73cm}C{0.9cm}C{0.73cm}C{0.78cm}C{0.5cm}}
\hline
                           & \multicolumn{2}{c}{Disturbed}              & \multicolumn{2}{c}{\makecell{Not dist.}}              & \multicolumn{2}{c}{Uncertain}             \\
                           & AGN                        & Cont.        & AGN                                        & Cont.        & AGN                        & Cont.       \\ \hline
\makecell{RI-HERG\\low}    & 37 $^{+9}_{-8}$             & 24 $^{+4}_{-3}$ & 53 $\pm$ 9                               & 69 $^{+3}_{-4}$ & 10 $^{+8}_{-3}$               & 7 $^{+3}_{-2}$ \\

\makecell{RI-HERG\\high}   & 57 $^{+9}_{-10}$               & 35 $^{+5}_{-4}$ & 39 $^{+10}_{-8}$                               & 60 $^{+4}_{-5}$ & 4 $^{+7}_{-1}$                & 6 $^{+3}_{-1}$ \\

\makecell{3CR\\ }     & 53 $\pm$ 6               & 27 $^{+3}_{-2}$ & 39 $^{+6}_{-5}$                               & 67 $\pm$ 3 & 8 $^{+4}_{-2}$               & 7 $^{+2}_{-1}$ \\

\makecell{\textit{-- 3CR}\\\textit{HERGs}}      & 66 $^{+7}_{-8}$              & 27 $^{+4}_{-3}$ & 24 $^{+8}_{-5}$                               & 66 $^{+3}_{-4}$ & 10 $^{+7}_{-3}$               & 7 $^{+2}_{-1}$ \\

\makecell{\textit{-- 3CR}\\ \textit{LERGs}}      & 37 $^{+9}_{-8}$               & 26 $^{+4}_{-3}$ & 57 $^{+8}_{-9}$                               & 68 $\pm$ 4 & 7 $^{+8}_{-2}$                & 6 $^{+3}_{-1}$ \\

\makecell{Type 2\\quasars} & 64 $^{+8}_{-10}$              & 26 $\pm$ 4 & 36 $^{+10}_{-8}$                              & 69 $\pm$ 4 & --                          & 6 $^{+3}_{-1}$ \\ \hline
\end{tabular}
\end{table}

\begin{figure*}
    \centering
	\includegraphics[width=0.95\linewidth]{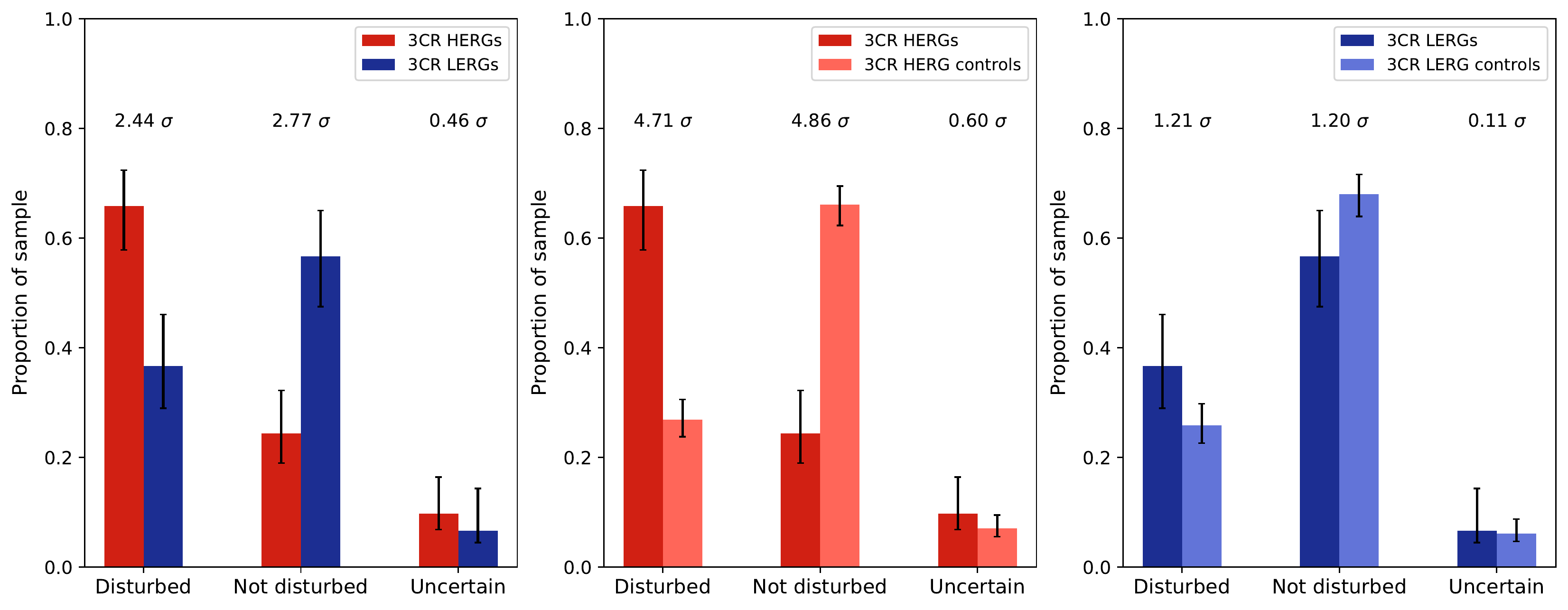}
    \caption[The proportions of 3CR HERGs, 3CR LERGs, and their respective matched control galaxies classed as disturbed, not disturbed or uncertain using the online interface.]{As Figure~\ref{fig:q1_all_agn_samples}, but for the HERGs and LERGs in the 3CR sample. In this case, the results for the two samples are presented both alongside each other (first panel) and with those for their respective matched control samples (second and third panels).}
    \label{fig:q1_all_HERGs_and_LERGs}
\end{figure*}

From these results, it is seen that the AGN show a preference for disturbed galaxies relative to the matched controls in all cases, supported by the measured proportions for both the disturbed and not disturbed categories. Across the radio AGN samples, the degree of significance for these differences appears to decrease with decreasing radio power, with the most significant excess being found for the 3CR sample ($>$4\,$\sigma$). The Type 2 quasars, however, also show a significant preference for disturbed morphologies relative to their matched controls (3.7\,$\sigma$), suggesting that the optical emission-line luminosity could also be important in this context. In-depth analysis of the relationships with radio power and optical emission-line luminosity is presented in the following subsection.

Previous study of the powerful radio galaxies in the 2Jy sample suggests that the hosts of radiatively-efficient radio AGN (the SLRGs) are typically more likely to be merging/interacting than those of radiatively-inefficient radio AGN \citep[the WLRGs;][]{ram11,ram12,ram13}. In terms of their rates of disturbance, \cite{ram11} found that 94 $^{+4}_{-7}$ per cent of the SLRGs and 27 $^{+16}_{-9}$ per cent of the WLRGs in the sample showed clear signatures of mergers and interactions, which differ at the 4.7\,$\sigma$ level according to a two-proportion Z-test. The inclusion of both HERGs and LERGs in the 3CR sample allows this picture to be tested with a larger sample.

Figure~\ref{fig:q1_all_HERGs_and_LERGs} again shows the measured proportions for the disturbed, not disturbed and uncertain categories outlined above, but in this case comparing the results for the 3CR HERG and LERG subsamples and their respective matched control samples (proportions also listed in Table~\ref{tab:q1_props}). These measurements appear to support the picture suggested by the 2Jy results, with HERGs showing an increased preference for disturbed morphologies both relative to their matched controls and to the LERGs. There is also evidence to suggest that the large difference between the disturbance rates for the 3CR sources and their matched control samples (seen in Figure~\ref{fig:q1_all_agn_samples}) is predominantly driven by the HERGs, with the HERG proportions showing $\sim$5\,$\sigma$ differences with respect to those of their matched controls and the LERG proportions only exhibiting $\sim$1\,$\sigma$ differences. 

However, the significance of the difference between the HERG and LERG proportions is lower than that found between the 2Jy SLRGs and WLRGs, with the two-proportion Z-test suggesting that the null hypothesis of them being the same can only be rejected at the 2.4\,$\sigma$ level. One caveat with this comparison is that, as mentioned \S\ref{sec:intro}, the SLRG/WLRG and HERG/LERG classification schemes are not completely equivalent \citep[despite considerable overlap;][]{tad16}. The 2Jy classifications were also obtained through more detailed morphological analysis based on higher quality imaging observations. The effects of these factors on this comparison are discussed in more detail in \S\ref{sec:disc}.

Given that the proportions of galaxies in each sample that were classified as uncertain were small, the trends in the results found for the disturbed galaxies are largely consistent with the opposite trends found for those classified as not disturbed, as can be seen in Figures~\ref{fig:q1_all_agn_samples} and \ref{fig:q1_all_HERGs_and_LERGs}. Therefore, only the galaxies securely classified as disturbed (i.e. above the five-vote threshold) are considered for the remainder of the analysis in this subsection.

\subsubsection{Relationship with radio power and [OIII]$\lambda$5007 luminosity}
\label{subsubsec:q1-RP_and_LOIII}

The detailed morphological analysis of the RI-HERG low sample presented in Paper I suggested that the association between AGN and merging galaxies could be strongly dependent on radio power, but more weakly dependent on optical emission-line luminosity. The significant excesses in the proportions of disturbed galaxies in the 3CR and Type 2 quasar samples, both relative to their respective matched control samples and to the RI-HERG samples, suggested that both properties could be important, but did not provide clear evidence as to which is the main driver of the trend. 

\begin{figure}
    \centering
	\includegraphics[width=\columnwidth]{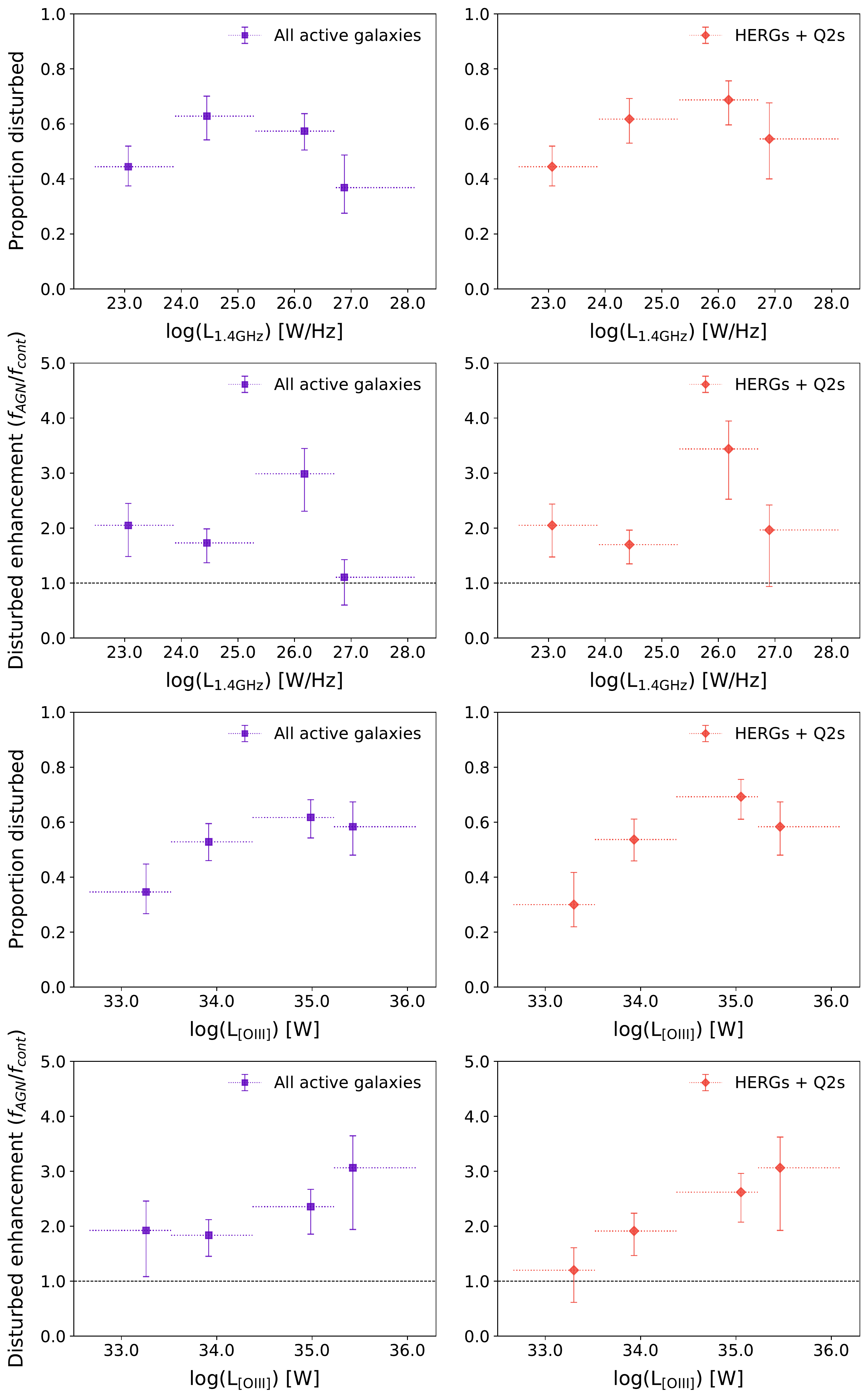}
    \caption{The disturbed proportions and enhancement ratios ($f_{\rm AGN}/f_{\rm cont}$) in bins of logarithmic 1.4 GHz radio power and [OIII]$\lambda$5007 emission-line luminosity for all active galaxies (left column) and for all HERGs and Type 2 quasars (right column) classified using the online interface. In all cases, the markers represent the median values for the active galaxies in each bin, the ranges of which are indicated by the dotted lines. A line representing equality between the proportions measured for active galaxies and their matched control galaxies is shown on the enhancement plots.}
    \label{fig:q1_dist_enh_vs_RP_OIII}
\end{figure}

The picture suggested by these results is complicated by the fact that there could be some underlying co-dependence between the AGN 1.4 GHz radio powers and [OIII]$\lambda$5007 luminosities. Pearson correlation tests\footnote{Active galaxies with upper limits on either their 1.4 GHz radio powers or [OIII]$\lambda$5007 luminosities were not considered for correlation tests or for the plots in Figure~\ref{fig:q1_dist_enh_vs_RP_OIII}.} suggest that there is a moderate but significant positive correlation between the two parameters in the RI-HERG high and 3CR samples ($r=0.517$, $p=0.005$ and $r=0.533$, $p<10^{-5}$, respectively), but that they are not significantly correlated in the RI-HERG low and Type 2 quasar samples ($r=0.077$, $p=0.648$ and $r=0.118$, $p=0.591$, respectively). This is consistent with the findings of previous studies, where optical emission-line luminosity and radio power are seen to be strongly correlated for powerful radio AGN, but more weakly correlated for those with lower radio powers \citep[e.g.][]{rs91,zb95,best05b}. A significant but weaker positive correlation is found when all of the active galaxy samples are combined ($r=0.315$, $p<10^{-4}$), although this is driven by the stronger correlations seen for the RI-HERG high and 3CR sources, which represent the majority of the total objects (100 out of 155). This suggests that AGN radio power and [OIII]$\lambda$5007 luminosity are not strongly correlated within the sample as a whole.

Figure~\ref{fig:q1_dist_enh_vs_RP_OIII} shows the proportions of galaxies classified as disturbed in bins of 1.4 GHz radio power and [OIII]$\lambda$5007 luminosity for all of the active galaxies in the current project. 
These proportions are also expressed as ``enhancements" relative to those measured for their matched controls, i.e. the ratios of the fractions of disturbed active galaxies in each bin to those found for the corresponding matched control galaxies ($f_{\rm AGN}/f_{\rm cont}$). The distributions for all active galaxies (including LERGs) and for all HERGs and Type 2 quasars are presented separately. 

Across the full range of radio powers covered, the distributions with 1.4 GHz radio power for the disturbed proportions and their enhancements are not significantly different for the full active galaxy sample and for the HERG and Type 2 quasar objects. Both are consistent with a general increase in the rates of disturbance in the two lower-radio-power bins, but the differences between the proportions in these two bins are not highly significant (at the 1.5\,$\sigma$ level), and this trend is not seen when the enhancement ratios are considered. While the enhancement ratios for both the full active galaxy sample and the HERG/Type 2 quasar subset reveal $\sim$3\,$\sigma$ excesses above equality between the active galaxy and control galaxy disturbance rates in the third radio power bin ($\rm 25.31 \leq log(L_{1.4GHz}) \leq 26.73$ W\,Hz$^{-1}$), no significant excess is observed at higher radio powers. 

As found for the distributions with radio power, the disturbed proportions for the full active galaxy sample and the HERG/Type 2 quasar subset are consistent over the full range of [OIII]$\lambda$5007 luminosity covered. Suggestions of a general increase towards higher luminosities are seen, but again have low significance. In this case, however, a positive, seemingly linear trend with [OIII]$\lambda$5007 luminosity is observed in the disturbed enhancement ratio distribution for the HERGs and Type 2 quasars. A Pearson correlation test provides evidence for a strong positive correlation between the disturbed enhancement ratios and the median [OIII]$\lambda$5007 luminosities for the binned HERG/Type 2 quasar data, significant at the 97.2 per cent level: $r_{\rm HERG}=0.972$, $p_{\rm HERG}=0.028$. Bootstrapping analysis shows that the strength of this correlation is not sensitive to the uncertainties in the measured enhancements.

A strong trend with [OIII]$\lambda$5007 luminosity is not seen in the disturbed enhancements for the full active galaxy sample, and a Pearson correlation test in this case indicates no highly significant evidence for a correlation: $r_{\rm full}=0.772$, $p_{\rm full}=0.228$. The relationship is therefore stronger when the LERGs are excluded, and only the HERGs and Type 2 quasar hosts are considered. This suggests that mergers and interactions become increasingly important for triggering radiatively-efficient AGN towards higher optical emission-line luminosities, but that the same relationship does not apply to radiatively-inefficient AGN.

Overall, the results provide evidence that the importance of galaxy mergers and interactions for triggering radiatively-efficient AGN is strongly dependent on [OIII]$\lambda$5007 emission-line luminosity, but not strongly dependent on 1.4 GHz radio power, in contrast with the results from the more detailed analysis of the RI-HERG low sample (Paper I). Given the relationship between [OIII]$\lambda$5007 luminosity and the total AGN power \citep[e.g.][]{heck04}, this supports the idea of an increasing importance of merger-based AGN triggering towards higher bolometric luminosities, as suggested by previous studies in the literature \citep[e.g.][]{tre12}. The observed difference between the distributions of the proportions of disturbed active galaxies and their enhancement ratios also serves to highlight the importance of the control matching process. 

\subsubsection{Relationship with stellar mass and redshift}
\label{subsubsec:q1-mass_z}

\begin{figure}
    \centering
	\includegraphics[width=0.85\linewidth]{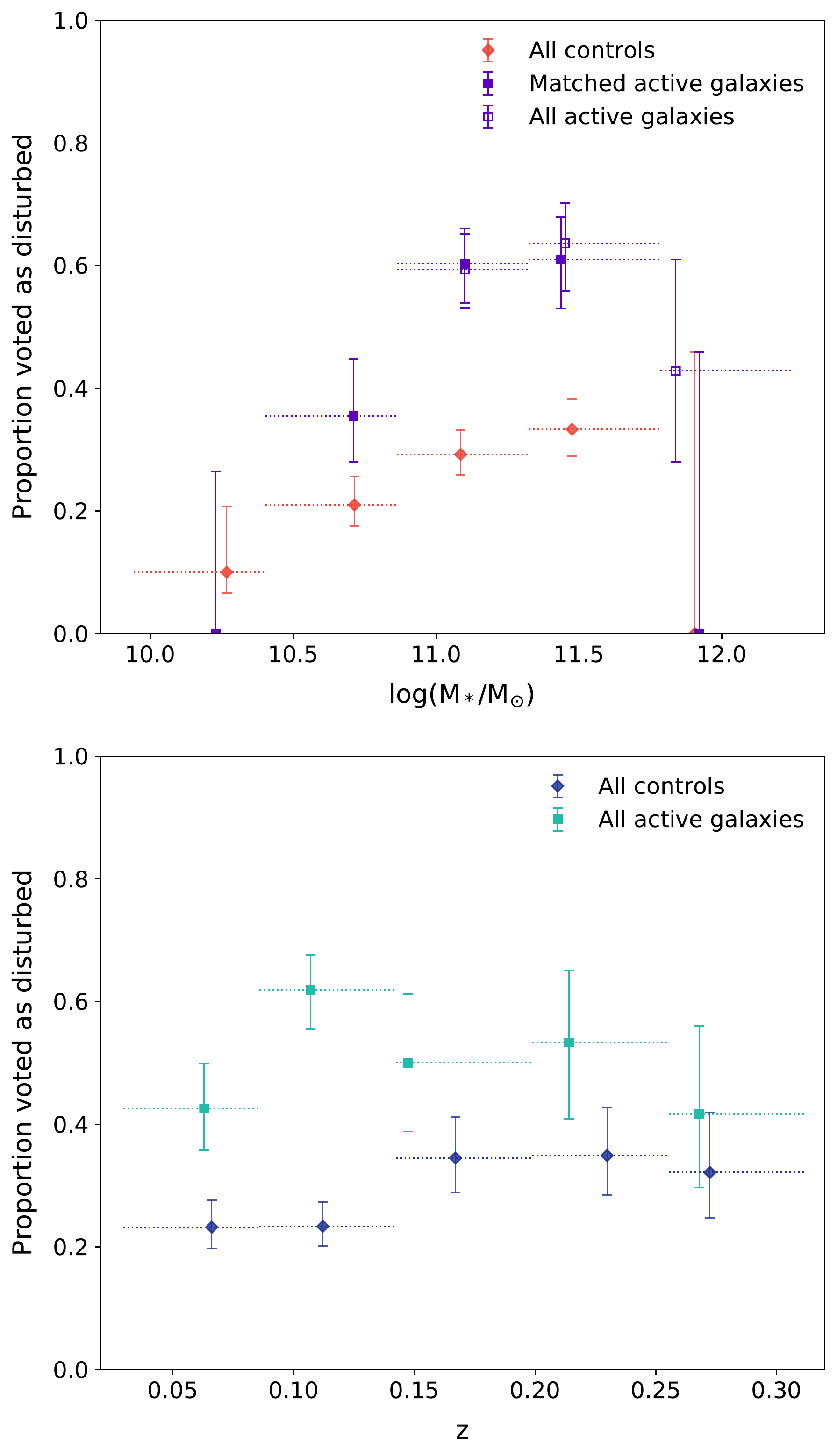}
    \caption[]{The proportions of active galaxies and matched control galaxies classified as disturbed in bins of stellar mass and redshift. The proportions are plotted at the median stellar masses and redshifts of the active galaxies and control galaxies in each bin. Proportion uncertainties were estimated following the method of \citet[][]{cam11}. For the plot against stellar mass, the results for the full active galaxy sample (unfilled points, including the unmatched 3CR sources) and the active galaxies with control matches (filled points) are shown separately.}
    \label{fig:q1_dist_frac_vs_mass_z}
\end{figure}

Investigation of the relationship between stellar mass and disturbance rate is important for ensuring that the particularly large stellar masses of the unmatched objects in the 3CR sample did not strongly affect the results outlined in the previous subsections. Figure~\ref{fig:q1_dist_frac_vs_mass_z} shows the distributions of disturbed proportions for the active galaxies and matched control galaxies with both stellar mass and redshift -- note that 3C 130 was not considered for the former plot due to its abnormally large stellar mass relative to the other galaxies in the project (log($\rm M_{*}/M_{\odot}$) = 12.7). The distributions with stellar mass for the matched active galaxies and the full active galaxy sample are shown separately, to illustrate the effect of including the unmatched 3CR objects. The distributions with redshift for these two samples are very similar, and so only the results for the full active galaxy sample are shown, for clarity.

From Figure~\ref{fig:q1_dist_frac_vs_mass_z}, it is seen that the active galaxies have higher rates of disturbance than the matched controls across the full range of stellar mass and redshift, except at the lowest and highest stellar masses. This confirms that the active galaxies are in general more frequently disturbed, as originally indicated by the proportions presented in Table~\ref{tab:q1_props} and Figure~\ref{fig:q1_all_agn_samples}.

No clear trend is visible with redshift for either the active galaxies or matched controls. The highly significant excess in the disturbed fraction for the active galaxies with $0.08 \lesssim z \lesssim0.14$ (4.6\,$\sigma$) is likely caused by the high rates of disturbance found for the RI-HERG high and Type 2 quasar samples, the median redshifts of which both lie within this bin ($z=0.110$ and $z=0.111$, respectively). On the other hand, a positive trend with stellar mass is seen for both the active galaxies and the matched control galaxies. This is steeper for the active galaxies, and therefore the significance of the excess in the disturbed proportions relative to the matched controls is seen to increase with stellar mass in this range: 5.0\,$\sigma$ and 4.3\,$\sigma$ in the third and fourth bins, respectively, for the full active galaxy sample. 
Due to the small numbers of active galaxies and control galaxies at the highest stellar masses, the uncertainties on the disturbance proportions are large. As a result, there is no strong evidence to suggest that the inclusion of the unmatched 3CR objects has a significant affect on any differences between active and control galaxy disturbance rates.

\subsection{Interaction signatures and merger stage}
\label{subsec:features_merger_stage}

The second question in the online interface asked the researchers to identify the specific types of interaction signature that they had seen in the galaxy images, with the goal of better characterising the types and stages of the mergers and interactions identified. Classifiers were asked to answer the question ``What types of interaction signature are visible?" using one or more of the categories outlined in \S\ref{subsec:interface}.

This question was only answered in cases where classifiers had already indicated that clear interaction signatures were visible in the galaxy images by responding ``Yes" to the first question. Multiple responses could be selected, since several different types of interaction signature could be present at the same time. As a result of these two factors, the total number of votes recorded for each galaxy across the different interaction signature categories varied. An interaction signature classification was therefore required to meet two criteria in order to be accepted. Firstly, the threshold of 5 out of 8 votes must have been met for the first question, ensuring that only the galaxies accepted as being disturbed were included in the analysis. Secondly, the majority of the classifiers that had answered ``Yes" to the first question must have voted for that particular interaction signature category, i.e. 3 out of 5, 4 out of 6, 4 out of 7, or 5 out of 8. 

The main aim of this question was to determine whether the galaxy appeared to be in the early stages or late stages of a merger or interaction, i.e. prior to or following the coalescence of the galaxy nuclei (``pre-coalescence" or ``coalescence/post-coalescence", respectively). For this purpose, the ``Bridge (B)", ``Tidally interacting companion (TIC)" and ``Multiple nuclei (MN)" classifications were considered to indicate early-stage or pre-coalescence interactions, while the remainder were classified as late-stage or coalescence/post-coalescence interactions (late-stage or post-coalescence hereafter, for brevity). This approach is mostly consistent with that of \cite{ram11,ram12} and \cite{bess12}, with the exception of the new ``Tidally interacting companion (TIC)" category that was added for this work. On the occasions when both early-stage and late-stage interaction signature classifications were accepted, the interaction was considered to be of early-stage/pre-coalescence, for consistency with these previous studies.

\begin{figure}
    \centering
	\includegraphics[width=0.93\columnwidth]{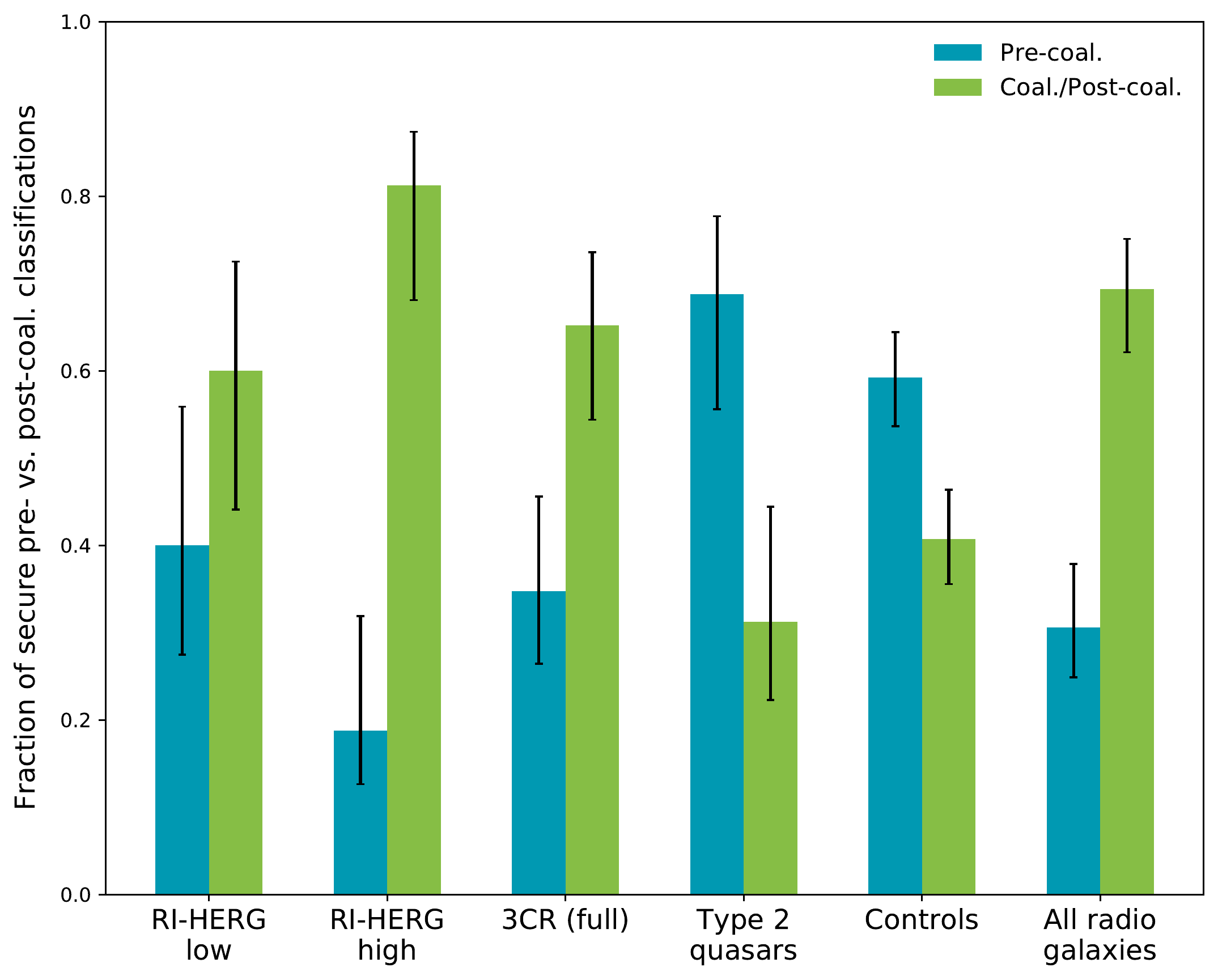}
    \caption{The proportions of disturbed galaxies in each of the active galaxy samples that show evidence for pre-coalescence or post-coalescence interactions. The values for the disturbed galaxies in the full control sample {and for all radio galaxy samples combined} are also presented. Only galaxies with secure classifications of pre- or post-coalescence interaction signatures were considered for the analysis (see text), and the proportions shown were derived relative to the combined total number of these classifications for each sample. The exact measured proportions are listed in Table~\ref{tab:q2_pre_vs_post_merger}.}
    \label{fig:q2_pre_vs_post_merger}
\end{figure}

Figure~\ref{fig:q2_pre_vs_post_merger} shows the proportions of disturbed radio galaxies, Type 2 quasar objects, and control galaxies with secure interaction signature classifications that are suggestive of early-stage/pre-coalescence or late-stage/post-coalescence events based on the categorisation outlined above. The measured proportions for each sample are provided in Table~\ref{tab:q2_pre_vs_post_merger}. 

Late-stage interactions appear to be slightly favoured for the disturbed galaxies in each radio galaxy sample, although the late-stage and early-stage proportions are consistent with being equal for the RI-HERG low sample. This preference is also found when considering the measured proportions for all radio galaxy samples combined: 31 $^{+7}_{-6}$ per cent and 69 $^{+6}_{-7}$ per cent for early- and late-stage interaction signatures, respectively. These results agree well with the those found for the powerful radio galaxies in the 2Jy sample, for which 35 $\pm$ 11 per cent and 65 $\pm$ 11 per cent of the disturbed objects with $z < 0.3$ show a preference for pre- and post-coalescence interactions, respectively \citep{ram11}.

\begin{table}
\centering
\small
\caption{The proportions of galaxies with interaction signatures that are indicative of pre-coalescence or post-coalescence mergers and interactions, as presented in Figure~\ref{fig:q2_pre_vs_post_merger}. All proportions are expressed relative to the combined total number of secure pre- and post-coalescence classifications within each sample, and are given as percentages. Proportion uncertainties were estimated following the method of \citet[][]{cam11}.} 
\label{tab:q2_pre_vs_post_merger}
\begin{tabular}{lcc}
\hline 
                           & Pre-coalescence & Post-coalescence \\ \hline
RI-HERG low  & 40 $^{+16}_{-13}$  & 60 $^{+13}_{-16}$    \\
RI-HERG high   & 19 $^{+13}_{-6}$  & 81 $^{+6}_{-13}$  \\
3CR (full)    & 35 $^{+13}_{-9}$   & 65 $^{+9}_{-13}$    \\
Type 2 quasars & 69 $^{+9}_{-13}$  & 31 $^{+13}_{-9}$   \\
All controls   & 59 $^{+5}_{-6}$  & 41 $^{+6}_{-5}$  \\
All radio galaxies  & 31 $^{+7}_{-6}$  & 69 $^{+6}_{-7}$   \\
 \hline
\end{tabular}
\end{table}

On the other hand, the Type 2 quasar hosts show a preference for early-stage ({69 $^{+8}_{-13}$ per cent}) relative to late-stage ({31 $^{+13}_{-8}$ per cent}) interactions. {These proportions are, however, consistent with those found for the disturbed galaxies in the control sample, for which 59 $^{+6}_{-7}$ per cent and 41 $^{+7}_{-6}$ per cent were found to have early-stage and late-stage interaction signatures, respectively.} Two-proportion Z-tests suggest that the null hypothesis that the proportions of early-stage (or late-stage) interactions in the control sample and Type 2 quasar sample are the same can only be rejected at a confidence level of {0.7\,$\sigma$}. {Similar fractions of pre-coalescence and post-coalescence interactions were also found for the disturbed hosts of Type 2 quasars with moderate redshifts ($0.3 < z < 0.41$) studied by \cite{bess12}: 47 $\pm$ 13 per cent and 53 $\pm$ 13 per cent, respectively.}

{The} measurements for the Type 2 quasar sample and control sample contrast significantly with those found for the radio galaxies, with the proportions of early-stage (or late-stage) interactions differing with those in the combined radio galaxy sample at confidence levels of $\sim$3\,$\sigma$, in both cases. {Note that all} of the above results are preserved if the radio galaxies with [OIII]$\lambda$5007 luminosities above the quasar-like threshold ($\rm L_{[OIII]} \geq 10^{35}$ W) are not considered for the analysis -- {the early-stage and late-stage proportions become 30 $^{+8}_{-6}$ per cent and 68 $^{+6}_{-8}$ per cent, respectively}.

Overall, the interaction signature classifications from the interface therefore {tentatively} suggest that radio AGN are preferentially triggered in the late stages of galaxy mergers and interactions, while Type 2 quasars are preferentially triggered in their early stages. 
Regardless of this, AGN host galaxies with both pre- and post-coalescence interaction signatures are found in each of the active galaxy samples. Therefore, if the galaxy mergers and interactions are responsible for triggering each of the types of AGN considered, this can occur at several different phases during these events. 

\subsection{Morphological types}
\label{subsec:morph_types}

The final question in the online interface asked the classifiers to indicate the overall morphological type of the galaxy in the image. Classifiers were required to answer the question ``On first impression, what is the morphological type of the galaxy?" by selecting spiral/disk, elliptical, lenticular, merger (too disturbed to categorise), or indicating that it was unclassifiable due to image issues (\S\ref{subsec:interface}). This was done to test if there was any variation in the longstanding relationship between radio AGN and early-type host galaxies with radio power, as suggested by the mixed population of late- and early-type galaxies found for the RI-HERG low sample in Paper I. 
All classifiers were required to answered this question, and only one of the available responses could be selected for each galaxy. As a result, a threshold of 5 out of 8 votes was again used for accepting a classification. If this was not met for any of the options (excluding the ``Unclassifiable (due to image defects, \textit{not} merger)" category), the galaxy was classed as having an uncertain host type.

\begin{table*}
\centering
\caption{The proportions of galaxies with host types classed as elliptical, spiral/disk, lenticular, merger (too disturbed to place in former categories), or uncertain for all of the active galaxy and matched control samples classified using the online interface. All proportions are expressed as percentages{, and the number of objects in each sample are also presented}. The results for the 3CR HERG and LERG subsamples are included in separate columns, alongside those found for the full 3CR sample. Proportion uncertainties were estimated following the method of \citet[][]{cam11}.} 
\label{tab:q3_props}
\begin{tabular}{cccccccccccc}
\hline 
&  & \multicolumn{2}{c}{Elliptical}           & \multicolumn{2}{c}{\makecell{Spiral/disk}}          & \multicolumn{2}{c}{Lenticular}        & \multicolumn{2}{c}{Merger}       & \multicolumn{2}{c}{Uncertain}              \\
      & { $N$}                   & AGN                            & Cont.        & AGN                                        & Cont.        & AGN                            & Cont.       & AGN                        & Cont.       & AGN                           & Cont.        \\ \hline
\makecell{RI-HERG low}  & { 30}  & 17 $^{+9}_{-5}$                   & 49 $\pm$ 4 & 47 $\pm$ 9                               & 34 $\pm$ 4 & 10 $^{+8}_{-3}$                  & 4 $^{+2}_{-1}$ & 0 $^{+6}$                         & 1 $^{+2}$ & 27 $^{+9}_{-6}$                  & 11 $^{+3}_{-2}$ \\
\makecell{RI-HERG high} & { 28} & 60 $^{+8}_{-9}$                  & 56 $^{+4}_{-5}$ & 14 $^{+9}_{-4}$                               & 25 $^{+4}_{-3}$ & 0 $^{+6}$                             & 3 $^{+2}_{-1}$ & 4 $^{+7}_{-1}$                & 2 $^{+2}_{-1}$ & 21 $^{+10}_{-6}$                  & 14 $^{+4}_{-3}$ \\
\makecell{3CR (full)}  & { 72}   & 86 $^{+3}_{-5}$                   & 73 $^{+2}_{-3}$ & 1 $^{+3}$                               & 16 $\pm$ 2 & 0 $^{+2}$                             & 1 $^{+1}$ & 3 $^{+3}_{-1}$                & 1 $^{+1}$ & 10 $^{+5}_{-2}$                  & 9 $\pm$ 2  \\
\makecell{3CR HERGs}  &  { 41}   & 78 $^{+5}_{-8}$                   & 78 $\pm$ 3 & 2 $^{+5}_{-1}$                                & 14 $^{+3}_{-2}$ & 0 $^{+4}$                             & 2 $^{+2}_{-1}$ & 5 $^{+6}_{-2}$                & 1 $^{+1}$ & 15 $^{+7}_{-4}$                  & 6 $^{+2}_{-1}$  \\
\makecell{3CR LERGs}  & { 30}    & 97 $^{+1}_{-7}$                   & 71 $^{+3}_{-4}$ & 0 $^{+6}$                                         & 16 $^{+4}_{-3}$ & 0 $^{+6}$                             & 1 $^{+2}$ & 0 $^{+6}$                         & 1 $^{+2}$ & 3 $^{+7}_{-1}$                   & 11 $^{+3}_{-2}$ \\
\makecell{Type 2 quasars} & { 25} & 52 $^{+9}_{-10}$                & 47 $^{+5}_{-4}$ & 8 $^{+9}_{-3}$                                & 37 $^{+5}_{-4}$ & 4 $^{+8}_{-1}$                    & 5 $^{+3}_{-1}$ & 12 $^{+10}_{-4}$   &  1 $^{+2}$  & 24$^{+10}_{-6}$   &  9 $^{+3}_{-2}$   \\
 \hline
\end{tabular}
\end{table*}

The measured morphological type proportions for all samples are provided in Table~\ref{tab:q3_props}. With the exception of the RI-HERG low sample (27 $^{+9}_{-6}$ per cent early type, 47 $\pm$ 9 per cent late type), it is found that the majority of the galaxies in each radio galaxy sample were classified as having early-type morphologies (elliptical and lenticular; dominated by the former), and small proportions were deemed to have late-type morphologies (spiral or disk-like). Their respective matched controls show relative excesses in late-type morphologies in all cases, with the same exception. The majority of the Type 2 quasar hosts are also classed as ellipticals, and a significant excess in the merger category relative to their matched controls is also found (3.1\,$\sigma$), consistent with the high overall rate of disturbance determined for the sample (64 $^{+8}_{-10}$ per cent; \S\ref{subsec:dist_rates}). 

In order to investigate the relationship between AGN host type and radio power, the proportions of active galaxies with morphologies classified as early-type and late-type were compared across the full range of radio powers covered. Here, all galaxies classified as either elliptical or lenticular were considered to be of early-type, and those classed as spirals or disks were considered to be of late-type, consistent with the analysis presented in Paper I. 

\begin{figure}
    \centering
	\includegraphics[width=0.98\columnwidth]{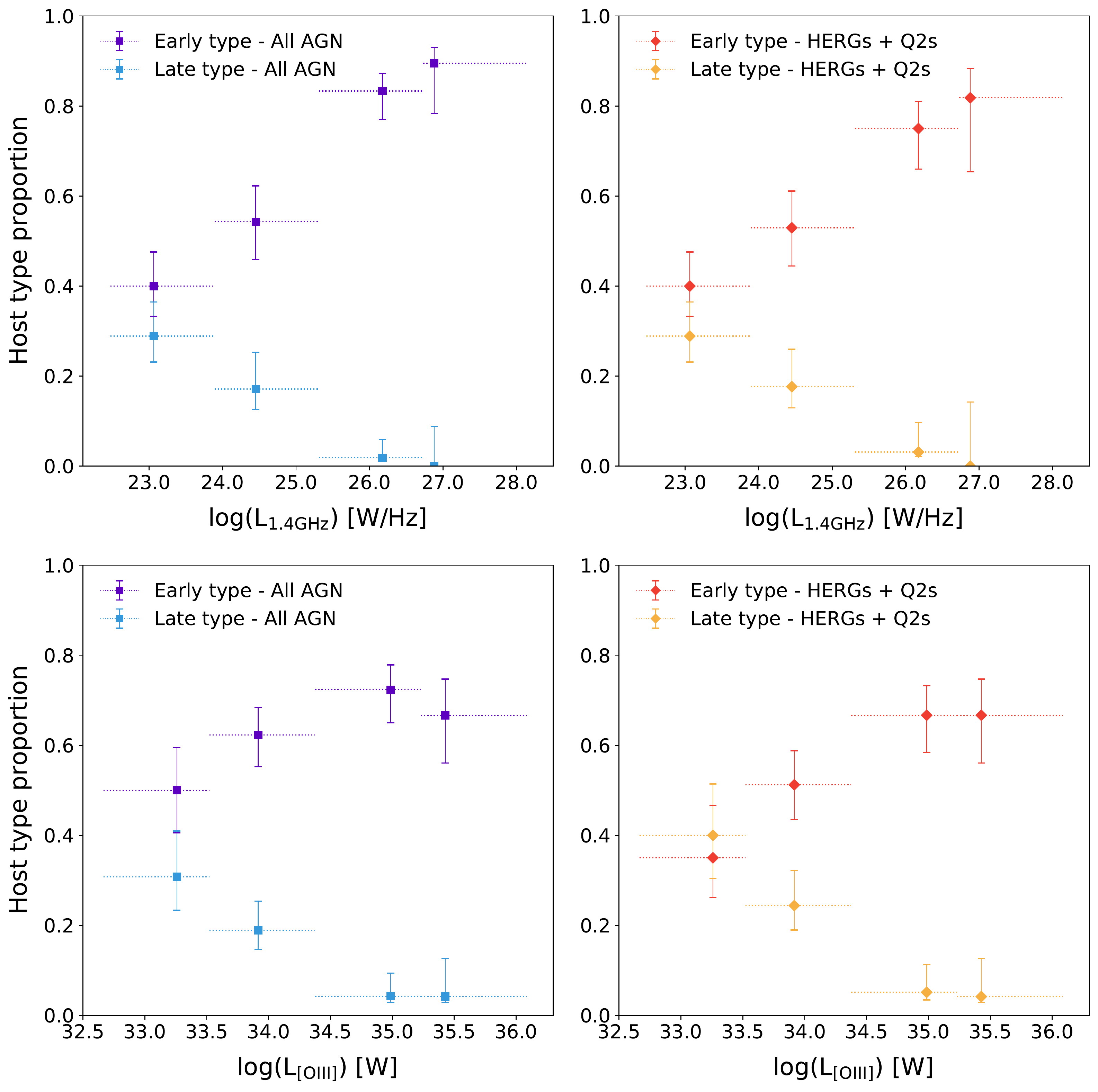}
    \caption{The proportions of active galaxies classed as having early-type (elliptical or lenticular) and late-type (spiral or disk) morphologies against their 1.4 GHz radio powers (top panels) and [OIII]$\lambda$5007 emission-line luminosities.}
    \label{fig:q3_host_types_vs_RP}
\end{figure}

Figure~\ref{fig:q3_host_types_vs_RP} shows the proportions of early-type and late-type galaxies in bins of 1.4 GHz radio power and [OIII]$\lambda$5007 emission-line luminosity for the full active galaxy sample and, separately, for only the HERGs and Type 2 quasars. A strong positive correlation is observed between the early-type proportions and medians of the 1.4 GHz radio power bins for both the full active galaxy sample ($r_{\rm full}=0.994$, $p_{\rm full}=0.006$) and the HERG and Type 2 quasar subset ($r_{\rm HERG}=0.998$, $p_{\rm HERG}=0.002$), according to Pearson correlation tests. This is coupled with a strong negative correlation for the proportion of late-type galaxies in both cases ($r_{\rm full}=-0.993$, $p_{\rm full}=0.007$ and $r_{\rm HERG}=-0.997$, $p_{\rm HERG}=0.003$). Pearson correlation tests also suggest strong correlations with [OIII]$\lambda$5007 luminosity, although these are found to be of lower significance than the relationships with 1.4 GHz radio power: $r_{\rm full}=0.868$, $p_{\rm full}=0.132$ and $r_{\rm HERG}=0.972$, $p_{\rm HERG}=0.028$, and $r_{\rm full}=-0.978$, $p_{\rm full}=0.022$ and $r_{\rm HERG}=-0.985$, $p_{\rm HERG}=0.003$ for the early- and late-type proportions, respectively. In contrast with the rates of disturbance, this suggests that the host types are more strongly linked with the radio power of the AGN than the optical emission-line luminosity.

While the control galaxies were matched to the targets in terms of their stellar masses and redshifts, matching of their morphological types was not performed. It is therefore important to check that the difference between the rates of disturbance for the active galaxies and control galaxies were not caused by the general preference for late-type morphologies exhibited by the latter objects (Table~\ref{tab:q3_props}).
Across the full sample of galaxies studied in the project, including all active galaxies and control galaxies, it is found that 30 $^{+3}_{-2}$ per cent of the early types and 26 $\pm$ 4 per cent of the late types were classified as disturbed. A two-proportion Z-test indicates that the null hypothesis that these two proportions are the same can be rejected at the 0.9\,$\sigma$ level. There is thus little evidence for a significant difference between the disturbance rates of the early-type and late-type galaxies studied, and the general excess of late-type galaxies in the control samples relative to the active galaxy samples should therefore not affect the comparison of their disturbance fractions. Matching of the morphological types would hence have had little effect on the results presented in \S\ref{subsec:dist_rates} \citep[as found by][]{gord19}.

Taken together, the results described in this section are consistent with the longstanding idea that the most powerful radio AGN are hosted by massive early-type galaxies. In addition, the fraction of early-type hosts is found to decrease strongly with decreasing radio power, while the proportion of late-type hosts shows the opposite trend. Since the sample is dominated by HERGs and Type 2 quasar objects, this supports the picture of a transition in the dominant host types for radiatively-efficient AGN from massive early-type galaxies at high radio powers to late-type hosts at low radio powers (i.e. like Seyfert galaxies), supporting the results from Paper I.

\section{Discussion}
\label{sec:disc}

\subsection{Comparison with more detailed inspection -- The RI-HERG low sample}
\label{subsec:method_disc}

The online interface provided a means for obtaining morphological classifications of a large sample of active galaxies and matched control galaxies in a time-efficient manner. The standardised image format, set classification questions/categories and the randomisation of the galaxy image presentation also allowed the levels of individual classifier bias to be greatly reduced. 
Prior to comparing with results from the literature, however, we here consider how classifications obtained for the RI-HERG low sample using the online interface compare with those obtained from the more detailed visual inspection (i.e. with the ability to manipulate the image contrast and scale as required) in Paper I.

When considering the online interface classifications, it is found that 11 out of 30 (37 $^{+9}_{-8}$ per cent) of the galaxies in the RI-HERG low sample are classed as disturbed, compared with the 16 out of 30 (53 $\pm$ 9 per cent) found from the more detailed analysis. The results from the two methods therefore show good overall agreement, with the same classifications (either both disturbed or both not disturbed) being determined for 77 per cent of the RI-HERG low objects (23 out of 30). Most importantly, this includes the eight galaxies that in Paper I were identified as ``highly disturbed" based on cursory visual inspection, all of which were classified as disturbed by either 7 (J1351+46, J1358+17) or all 8 (J0757+39, J0836+44, J0902+52, J1243+37, J1257+51, J1412+24) of the researchers in the online interface. 

Of the remainder, 3 galaxies (10 per cent of the sample) were classified as having an uncertain level of disturbance (4 votes disturbed, 4 votes not disturbed) through the online interface but as disturbed in the more detailed analysis (J0827+12, J1601+43, J1609+13). A further 3 galaxies (10 per cent) had secure classifications as not disturbed from the interface (i.e. meeting the 5-vote threshold) that disagreed with the disturbed classifications from the more detailed analysis (J0725+43, J0911+45, J1236+40). As can be seen from the images presented in Paper I, all of these galaxies exhibit subtle morphological signatures of disturbance, and, because of the limited image manipulation afforded by the interface method, the lack of agreement between the two methods is therefore unsurprising. The final galaxy, J1324+27, was the only object classified as disturbed when using the online interface but as not disturbed in the detailed analysis. This galaxy was a borderline case, however, with 5 votes recorded for disturbed and 3 for not disturbed, and it is seen to exhibit an unusual spiral structure that could be interpreted either as a sign of disturbance or as that of an undisturbed late-type galaxy.

When dividing the sample into two halves by radio power, the two methods give the same proportion of disturbed galaxies for the half with the highest radio powers (10 out of 15; 67 $^{+10}_{-13}$ per cent), but a much reduced proportion is found for the lower-radio-power half from the interface classifications relative to the detailed inspection -- 1 out of 15 (7 $^{+13}_{-2}$ per cent) and 6 out of 15 (40 $^{+11}_{-13}$ per cent), respectively. This suggests that the galaxies in the higher-radio-power half of the sample exhibit higher levels of disturbance than those in the lower-radio-power half, in support of the conclusions drawn based on the ``highly disturbed" galaxies in the sample in Paper I. From the interface classifications, the two-proportion Z-test now indicates that the null hypothesis that the two proportions are equal can be rejected at a confidence level of 3.4\,$\sigma$, compared to the value of 1.5\,$\sigma$ obtained previously. Repeating this analysis in terms of [OIII]$\lambda$5007 luminosity, proportions of 7 out of 15 (47 $\pm$ 17 per cent) and 4 out of 15 (27 $\pm$ 11 per cent) are measured for the high-luminosity and low-luminosity halves of the sample, respectively, a difference only at the 1.1\,$\sigma$ level. 

These results hence appear to support the idea that the importance of mergers and interactions for triggering radio AGN is strongly dependent on radio power but more weakly dependent on optical emission-line luminosity, as suggested in Paper I. However, they are based on the measured proportions of disturbed galaxies, which, as shown in \S\ref{subsec:dist_rates}, increase strongly with stellar mass for both active and non-active galaxies. Within the RI-HERG low sample, Pearson correlation tests suggest that there is a moderate but significant correlation between stellar mass and 1.4 GHz radio power ($r=0.569$, $p=0.001$), but no significant correlation between stellar mass and [OIII]$\lambda$5007 luminosity ($r=0.095$, $p=0.617$). While the disturbed proportions for radio-intermediate active galaxies are also positively correlated with radio power in the current project, this is not observed when the matched control galaxy proportions are taken into consideration (i.e. the enhancement ratios in Figure~\ref{fig:q1_dist_enh_vs_RP_OIII}). These factors therefore suggest that the apparent relationship with radio power {from the Paper I results} is in fact a consequence of an underlying trend with stellar mass in the general galaxy population, showing the importance of performing the control matching.

Considering the results for the host types, it is found that the classifications from the two methods agree for 18 of the 30 galaxies in the sample (60 per cent). However, 8 of the remainder were classed as uncertain (27 per cent), and so secure classifications (with $\geq$5 votes) from the interface only disagreed for 4 of the galaxies (13 per cent). Of this latter group, 3 out of 4 galaxies were classed as lenticular by one of the two methods and as late-type (spirals/disks) by the other, a difference that could be caused by the reduced ability to identify finer structures when using the interface. Interestingly, in the more detailed analysis, the host types for 5 out of the 8 uncertain cases from the interface classifications were deemed too disturbed to classify (the ``Merger" class), which could explain this categorisation. Therefore, although the rate of complete consistency is lower than for the classifications of morphological disturbance, the general agreement still appears to be good between the two methods, given these factors.

Overall, this comparison suggests that the interface classifications provide good sensitivity to major levels of disturbance but not to minor levels, and the derived rates of disturbance should therefore be treated as lower limits. Furthermore, the lower sensitivity to more subtle morphological details could lead to the preferential classification of early-type hosts relative to late-types. However, any limitations introduced by the interface method should affect the active galaxies and their matched control samples equally, and so conclusions based on relative comparisons between the two should be secure in all cases. This again highlights the importance of the control matching process carried out for the current work.

\subsection{The rates of disturbance and AGN triggering}
\label{subsec:mergers_and_triggering_disc}

The importance of galaxy mergers and interactions for triggering AGN has been the subject of much debate. The most widely accepted model suggests that radiatively-efficient AGN (e.g. HERGs/SLRGs/quasars) and radiatively-inefficient AGN (e.g. LERGs/WLRGs) differ in their dominant triggering and fuelling mechanisms \citep[e.g.][]{hb14,yn14}. In this picture, radiatively-efficient AGN are fuelled by a high Eddington rate cold gas flow from a standard accretion disk \citep[c.f.][]{ss73}, and hence a sufficient supply of such gas must be available to the central SMBH in order to initiate and sustain this type of nuclear activity. The strong inflows of gas caused by the tidal forces associated with galaxy mergers and interactions \citep[e.g.][]{bh96,gabor16} therefore provide an attractive mechanism for triggering and fuelling the AGN in these objects. 

Radiatively-inefficient AGN are then thought to be fuelled by an optically thin, geometrically thick accretion flow of hotter gas at low Eddington rates \citep[c.f.][]{ny94,ny95,nar05}. At high radio powers, the favoured fuelling mechanisms in this case are often linked with the prevalent hot gas supply in the host galaxy haloes and the dense larger-scale environments in which these active galaxies typically lie \citep[e.g.][]{baum92,best05b,hard07,gas13}, with mergers and interactions thus being relatively less important. The results obtained from the interface classification analysis are here discussed in this context.

\subsubsection{Powerful radio galaxies -- Comparison with the 2Jy sample}
\label{subsubsec:2jy_comp}

Previous deep, ground-based optical observations of powerful radio galaxies have revealed frequent morphological signatures of galaxy mergers and interactions, which, in keeping with the picture outlined above, are found to be more prevalent for those also exhibiting strong optical emission lines \citep{heck86,sh89a,sh89b,ram11}. {Furthermore, evidence from HST imaging observations suggests that the hosts of bright AGN with moderate to high radio luminosities also have high merger rates at higher redshifts \citep[$1<z<2.5$;][]{chi15}.}

\cite{ram11} found that 94 $^{+2}_{-7}$ per cent of the SLRGs in the 2Jy sample display these signatures, which were also found to preferentially lie in the moderate-density group environments that favour the frequent occurrence of these events \citep{ram13}. In contrast, evidence for morphological disturbance was found in only 27 $^{+16}_{-9}$ per cent of the 2Jy WLRGs \citep{ram11}, which were also found to be predominantly associated with denser cluster environments \citep{ram13}, where the high relative galaxy velocities can have a negative effect on the merger rate \citep{pb06}.
It was also found that the rates of disturbance are considerably lower for non-active early-type galaxies with comparable optical luminosities, redshifts and image depths than for the 2Jy SLRGs, showing disturbance fractions of 53 $\pm$ 7 per cent \citep[$z<0.2$, from the OBEY survey;][]{tal09} and 48 $\pm$ 5 per cent \citep[$0.2 \leq z < 0.7$, from the Extended Groth Strip;][]{zhao09} when interaction signatures with the same surface brightness limits were considered \citep[][]{ram12}. 

Considering the results obtained for the 3CR HERGs in the current analysis, it is seen that 66 $^{+7}_{-8}$ per cent of the objects are classed as disturbed based on the interface classifications, an excess at the 4.7\,$\sigma$ level relative to their matched controls. While the rate of disturbance is notably lower than the fraction determined for the 2Jy SLRGs at lower redshifts -- 93 $^{+2}_{-13}$ per cent \cite[13 out of 14 objects at $z<0.2$;][]{ram11} -- the excess relative to matched control galaxies in the OBEY survey for the latter is, in fact, less highly significant (at the 3.3\,$\sigma$ level). Both studies therefore suggest significant enhancements in disturbance rate for the hosts of radiatively-efficient AGN at high radio powers. The lower rates of disturbance found for both the 3CR HERGs and their matched controls are then likely accounted for by the reduced sensitivity to low surface brightness tidal features when using the interface method (\S\ref{subsec:method_disc}).

Turning to the 3CR LERGs, it is found that only 37 $^{+9}_{-8}$ per cent are classed as disturbed based on the interface classifications, which is consistent with the value of 20 $^{+17}_{-7}$ per cent determined for 2Jy WLRGs at the same redshifts (2 out of 10 objects at $z<0.2$). In addition, no significant evidence to suggest that the rates of disturbance differ from those of their matched controls is found, in both cases.

The results from the current analysis are therefore consistent with the idea that galaxy mergers and interactions are highly important for triggering the most powerful radio galaxies with radiatively-efficient AGN, but are much less important for triggering those with radiatively-inefficient AGN.

\subsubsection{Comparison with radio-intermediate LERGs}
\label{subsubsec:ri-lerg_comp}

Recently, study of the optical morphologies of a large sample of low redshift ($z<0.07$) LERGs with mostly intermediate radio powers ($\rm 10^{21.7} < log(L_{1.4GHz}) < 10^{25.8}$ W\,Hz$^{-1}$, median $10^{23}$ W\,Hz$^{-1}$) was undertaken by \cite{gord19}. The 282 LERGs were classified alongside 1622 control galaxies matched in stellar mass, redshift and large-scale environment, using a similar online interface technique to that used for the current analysis.  While the Dark Energy Camera Legacy Survey \citep[DECaLS;][]{dey19} images used by \cite{gord19} have a fainter limiting surface brightness depth \citep[$\mu_{r} \sim 28$ mag\,arcsec$^{-2}$;][]{hood18} than the INT/WFC images ($\mu_{r} \sim 27$ mag\,arcsec$^{-2}$), this is expected to have little effect on the classifications, given the reduced sensitivity to low-surface-brightness features when using the interface method (see \S\ref{subsec:method_disc}). These results are therefore directly comparable with those obtained from the online interface classifications of the current sample. The main caveat is that the classifiers were able to indicate whether the level of disturbance was ``major" or ``minor" when classifying the objects as disturbed, whereas the disturbed classifications in the current work encompass both.

The overall rates of disturbance determined for the radio-intermediate LERGs and their matched controls, considering classifications of both minor and major disturbances, are 28.7 $\pm$ 1.1 per cent and 27.3 $\pm$ 0.5 per cent, respectively, a difference at a confidence level of $<0.5$\,$\sigma$ \citep{gord19}. Considering the combined interface classification results for the RI-HERG low and RI-HERG high samples from the current analysis ($\rm 10^{22.5} < log(L_{1.4GHz}) < 10^{25}$ W\,Hz$^{-1}$; $z < 0.15$), disturbance rates of 47 $^{+7}_{-6}$ per cent and 29 $\pm$ 3 per cent are found for the radio-intermediate HERGs and their matched controls, a difference at the 2.7\,$\sigma$ level. Since the proportions measured for the control samples from both analyses are consistent ($<0.5$\,$\sigma$ difference), this suggests that the merger rates for radio-intermediate HERGs are significantly higher than for radio-intermediate LERGs -- a two-proportion Z-test indicates that the null hypothesis that the disturbance proportions are equal can be rejected at a confidence level of 2.7\,$\sigma$. 

In comparison, the rates of disturbance for the high-radio-power 3CR LERGs (37 $^{+9}_{-8}$ per cent) are consistent with both the radio-intermediate LERGs and the control sample. 
Therefore, while the 3CR LERGs have typically higher 1.4 GHz radio powers and [OIII]$\lambda$5007 emission-line luminosities\footnote{Two-sample KS tests suggests that the null hypothesis that the two LERG samples are drawn from the same underlying 1.4 GHz radio power and [OIII]$\lambda$5007 emission-line luminosity distributions can be rejected with very high confidence: $D = 0.972$, $p < 10^{-21}$ and $D = 0.852$, $p < 10^{-14}$ (upper limits excluded), respectively.}, there is no strong evidence that the rate of disturbance is significantly increased in this population. In support of this, \citet{ell15} find that their sample of LERGs with predominantly intermediate radio powers shows no significant excess in close pairs and post merger signatures relative to non-active controls, when both host galaxy properties and environmental structure are accounted for.

Overall, the results from both studies are hence consistent with the idea that galaxy mergers and interactions are generally less important for triggering the nuclear activity in LERGs than in HERGs. In addition, there is little evidence to support the idea that the importance of this triggering mechanism is dependent on either [OIII]$\lambda$5007 emission-line luminosity or 1.4 GHz radio power for radiatively-inefficient radio AGN. 

\subsubsection{The importance of mergers for triggering quasars}
\label{subsubsec:quasar_comp}

{Models of galaxy mergers and interactions suggest that they offer an effective means for triggering and fuelling quasar activity \citep[e.g.][]{sanders88,dm05,hop08}. However, previous searches for morphological disturbance in the hosts of bright AGN selected in different wavebands have yielded mixed results, both at low to intermediate \citep[e.g.][]{dun03,floyd04,benn08,veil09,cis11,tre12,vill14,hong15,vill17} and higher redshifts \citep[e.g.][]{koc12,koc15,chi15,glik15,mech16,don18,mar19,shah20}.} 

{The high merger rate and significant enhancement relative to the matched controls measured for our Type 2 quasar hosts (64 $^{+8}_{-10}$ per cent and 3.7\,$\sigma$) supports the idea that mergers and interactions provide the dominant triggering mechanism for luminous AGN. While other studies of the morphologies of Type 2 quasar objects have found both low and high disturbance rates \citep[e.g.][]{greene09,bess12,wyl16,urb19,zhao19}, there is strong evidence to suggest that the surface brightness depth of the observations is particularly important in this context. This issue will be discussed in a subsequent paper that will focus on the importance of mergers for triggering quasar-like nuclear activity (Pierce et al. in prep.), in which the results obtained from further deep imaging observations of Type 2 quasars will also be presented.}

\subsection{Host types}
\label{subsec:host_type_disc}

Evidence from many past observational studies suggests that powerful radio AGN ($\rm L_{1.4GHz} \gtrsim 10^{25}$ W\,Hz$^{-1}$) are predominantly associated with massive elliptical galaxies \citep[e.g.][]{mms64,dun03,best05b}. A minority of the objects in the high-flux-density selected 3CR and 2Jy samples do however exhibit disk-like morphologies upon cursory visual inspection, with these typically being found towards the lower end of the radio power range covered \citep{tad16}. Some of these objects, in fact, have intermediate radio powers, which supports previous suggestions that late-type hosts could be more common in this range \citep[e.g.][]{sad14}. Furthermore, the results presented in Paper I showed that HERGs in the intermediate radio power range have a mixture of early-type and late-type morphologies. In combination, these results could suggest that the hosts of radiatively-efficient AGN move towards the predominantly late-type morphologies of Seyfert galaxies \citep[e.g.][]{adams77} at lower radio powers.

The general trends observed for the morphological type proportions in the current analysis provide strong support for this picture. Pearson correlation tests revealed evidence for strong positive correlations with radio power for the proportions of active galaxies classified as early-type in both the full active galaxy sample and the HERG and Type 2 quasar subset, which are coupled with strong decreases in the proportions classified as late-type with increasing radio power -- each of these trends are clearly demonstrated in Figure~\ref{fig:q3_host_types_vs_RP}. While similar trends are also seen with [OIII]$\lambda$5007 emission-line luminosity, these are found to be of lower significance, thus indicating that the host types are more strongly linked with AGN radio power. These results therefore suggest that there is a gradual transition in the dominant host types of radio AGN from early-type galaxies at high radio powers to late-type galaxies at lower radio powers, at least for radiatively-efficient objects. They are also consistent with the idea that secular triggering mechanisms related to galaxy disks \citep[e.g.][]{hq10,hb14} become increasingly important towards lower radio powers, as suggested in Paper I, or lower total AGN powers \citep[as in e.g.][]{tre12}.

Looking at the results for the current Type 2 quasar sample, it is found that the majority of host galaxies are classed as early-type (52 $^{+9}_{-10}$ per cent elliptical, 4 $^{+8}_{-1}$ per cent lenticular) and only 2 out of 25 (8 $^{+9}_{-3}$ per cent) and 3 out of 25 (12 $^{+10}_{-4}$ per cent) are classed as late-type or ``merger", respectively; the remaining 24 $^{+10}_{-6}$ per cent have uncertain host types. A preference for early-type hosts has also been found in other imaging studies of low to intermediate redshift Type 2 quasars, from both visual inspection (\citeauthor{urb19} \citeyear{urb19}; but see \citeauthor{zhao19} \citeyear{zhao19}) and detailed light profile fitting \citep[][]{greene09,wyl16}. The preference for early-type hosts is enhanced when the radio-loud quasar-like AGN in the 3CR sample (with $\rm L_{[OIII]} \geq 10^{35}$ W) are considered together with the Type 2 quasars (73 $^{+6}_{-8}$ per cent), in agreement with previous results that suggest both radio-loud and radio-quiet quasars are predominantly hosted by elliptical galaxies \citep{dun03}. The host type classifications determined using the interface therefore appear to show good general agreement with those from these previous studies, and lend favour to the idea that powerful quasar-like activity in the local universe is largely associated with massive, early-type galaxies.

\section{Summary and conclusions}
\label{sec:summary}

Investigating the mechanisms that trigger AGN is key for correctly implementing their associated feedback processes in current models of galaxy evolution. The jets of radio AGN could be particularly important in this context, although little investigation of the dominant triggering and fuelling mechanisms for the lower radio power sources that comprise the bulk of the local population has been performed.

The morphologies of the radio-intermediate HERGs studied in the first paper in this series \citep[][]{pierce19} suggested that the importance of merger-based AGN triggering was strongly dependent on the radio power associated with the nuclear activity. However, there was some evidence to suggest that the optical emission-line luminosity may also play a role. Using an online classification interface, this paper has expanded the morphological analysis to a much larger sample of active galaxies that encompasses a broad range in both 1.4 GHz radio power and [OIII]$\lambda$5007 emission-line luminosity, allowing the dependence of AGN triggering by galaxy mergers and interactions on these properties to be investigated in more detail. The dependence of host galaxy type on the AGN radio power has also been assessed. This analysis has also been performed for large samples of control galaxies matched to the active galaxies in terms of both stellar mass and redshift, classified blindly alongside the active galaxies in a randomised manner. The main results are as follows.

\begin{itemize}
    \item The active galaxies are found to be more frequently disturbed than the matched control galaxies across the full range of stellar masses and redshifts covered by the samples. The most significant excesses are found for the 3CR (4.3\,$\sigma$) and Type 2 quasar samples (3.7\,$\sigma$). In the former case, this is largely driven by the HERGs in the sample, which show a 4.7\,$\sigma$ excess relative to their matched controls. The 3CR LERGs, by comparison, only show a 1.2\,$\sigma$ excess in their disturbance fraction. 
    \item There is no strong evidence that the rates of disturbance in the active galaxies are correlated with 1.4 GHz radio power when the rates measured for their matched controls are accounted for. In contrast, we find clear evidence that the enhancement in the disturbance rate for the HERGs and Type 2 quasar hosts relative to that of the matched controls ($f_{\rm AGN}/f_{\rm cont}$) increases strongly with [OIII]$\lambda$5007 luminosity: $r=0.972$, $p=0.028$, from a Pearson correlation test. A significant correlation is not found when the 3CR LERGs are included, suggesting that this relation applies only to radiatively-efficient AGN. 
    \item {The disturbed radio galaxies show a preference for post-coalescence interaction signatures relative to pre-coalescence signatures, which may suggest that these objects are more likely to be triggered in the later stages of galaxy mergers.} The Type 2 quasar hosts show the opposite preference, indicating that they are more likely to be triggered in the early stages.
    \item The AGN in all samples show a preference for early-type rather than late-type host galaxies, with the exception of the RI-HERG low sample (from Paper I). The latter sample exhibits a preference for late-type hosts and a significant deficit ($>3$\,$\sigma$) of early-types relative to its matched controls. The measured morphological type proportions also suggest that the fraction of early-type hosts decreases strongly with radio power, while the fraction of late-type hosts increases. This supports the idea that the dominant host types of radiatively-efficient radio AGN change from early-type galaxies at high radio powers to late-type galaxies at lower radio powers, as suggested in Paper I. This could also suggest that triggering via secular processes in galaxy disks holds more importance for the latter objects.
\end{itemize}

Overall, the measured rates of disturbance imply that the importance of galaxy mergers and interactions for triggering radiatively-efficient AGN (HERGs/Type 2 quasars) is strongly dependent on their optical emission-line luminosities (and hence bolometric luminosities) but not on their radio powers, once the disturbance rate in the underlying galaxy population is accounted for. Moreover, there is particularly strong evidence to suggest that galaxy mergers and interactions provide the dominant triggering mechanism for quasar-like AGN at low-to-intermediate redshifts, regardless of radio power. In contrast, these processes appear to be of much lower importance for triggering radiatively-inefficient radio AGN, since the majority of 3CR LERGs are associated with undisturbed elliptical galaxies. 

\section*{Acknowledgements}

The authors thank Mischa Schirmer for advice and assistance concerning image processing with \texttt{THELI}. YG and CO are supported by the National Sciences and Engineering Research Council of Canada (NSERC). CRA acknowledges financial support from the Spanish Ministry of Science, Innovation and Universities (MCIU) under grant with reference RYC-2014-15779, from the European Union's Horizon 2020 research and innovation programme under Marie Sk\l odowska-Curie grant agreement No 860744 (BiD4BESt), from the State Research Agency (AEI-MCINN) of the Spanish MCIU under grants ``Feeding and feedback in active galaxies" with reference PID2019-106027GB-C4 and ``Quantifying the impact of quasar feedback on galaxy evolution (QSOFEED)" with reference EUR2020-112266. CRA also acknowledges support from the Consejería de Econom\'{i}a, Conocimiento y Empleo del Gobierno de Canarias and the European Regional Development Fund (ERDF) under grant with reference ProID2020010105 and from IAC project P/301404, financed by the Ministry of Science and Innovation, through the State Budget and by the Canary Islands Department of Economy, Knowledge and Employment, through the Regional Budget of the Autonomous Community. {PSB acknowledges financial support from the State Research Agency (AEI-MCINN) and from the Spanish MCIU under grant "Feeding and feedback in active galaxies" with reference PID2019-106027GB-C42.}

\section*{Data Availability}


The postage stamp images used for the morphological classifications are available online as supplementary material, along with the full morphological classification results obtained for the active galaxies using the online interface. Additional data underlying this article will be shared on reasonable request to the corresponding author.



\bibliographystyle{mnras}
\bibliography{ref_master} 




\appendix








\bsp	
\label{lastpage}
\end{document}